\documentclass[3p,amsmath,amssymb,compress]{elsarticle}
\usepackage{amsmath}
\usepackage{graphicx}
\usepackage{latexsym}
\usepackage{amsmath,amssymb}
\usepackage{bm} 
\usepackage{xcolor}
\usepackage{stackrel}
\usepackage{accents}

\journal{Annals of Physics}

\numberwithin{equation}{section}

\newcommand{\St}{\mathop{\rm St}\nolimits}
\newcommand{\lgu}{l_{\mathrm{G}}}

\begin{document}
\begin{frontmatter}	

\title{Transport properties of strongly coupled electron-phonon liquids}
\author[WISC]{Alex Levchenko}
\author[KIT1,KIT2]{J\"{o}rg Schmalian}
\address[WISC]{Department of Physics, University of Wisconsin--Madison, Madison, Wisconsin 53706, USA}
\address[KIT1]{Institut f\"ur Theorie der Kondensierten Materie, Karlsruher Institut f\"ur Technologie, D-76131 Karlsruhe, Germany}
\address[KIT2]{Institut f\"ur Quantenmaterialien und Technologien, Karlsruher Institut f\"ur Technologie, D-76021 Karlsruhe, Germany}

\begin{abstract}
In this work we consider the hydrodynamic behavior of a coupled electron-phonon fluid, focusing on electronic transport under the conditions of strong phonon drag. This regime occurs when the rate of phonon equilibration due to e.g. umklapp scattering is much slower than the rate of normal electron-phonon collisions. Then phonons and electrons form a coupled out-of-equilibrium state where the total quasi-momentum of  the electron-phonon fluid is conserved.  A joint flow-velocity emerges as a collective hydrodynamic variable. We derive the equation of motion for this fluid from the underlying microscopic kinetic theory and elucidate its effective viscosity and thermal conductivity. In particular, we derive decay times of arbitrary harmonics of the distribution function and reveal its corresponding super-diffusive relaxation on the Fermi surface. We further consider several applications of this theory to magneto-transport properties in the Hall-bar and Corbino-disk geometries, relevant to experiments. In our analysis we allow for general boundary conditions that cover the crossover from no-slip to no-stress flows. Our approach also covers a crossover from the Stokes to the Ohmic regime under the conditions of the Gurzhi effect. In addition, we consider the frequency dependence of the surface impedance and non-equilibrium noise. For the latter, we notice that in the diffusive regime, a Fokker-Planck approximation,  applied to the electron-phonon collision integral in the Eliashberg form, reduces it to a differential operator with Burgers type nonlinearity. As a result, the non-equilibrium distribution function has a shock-wave structure in the energy domain. The consequence of this behavior for the Fano factor  of the noise is investigated. In conclusion we discuss connections and limitations of our results in the context of recent electron-phonon drag measurements in Dirac and Weyl semimetals, and layout directions for further extensions and developments.      
\end{abstract}

\date{\today}


\end{frontmatter}	

\tableofcontents

\section{Introduction and motivation} 

Hydrodynamic effects of electronic transport in quantum materials are of significant current interest in condensed matter physics; see reviews \cite{NGMS-HydroReview,Lucas-Fong-HydroReview}
and references therein. Various transport measurements in electrostatically defined wires in the two-dimensional electron gas in Ga(Al)As hetero-structures \cite{Molenkamp,Jong,Predel,Gao,Manfra}, monolayer and bilayer graphene \cite{Bandurin-1,Crossno,Ghahari,Kumar,Morpurgo,Bandurin-2,Berdyugin,Hone}, quasi-two-dimensional delafossite metals PdCoO$_2$ and PtCoO$_2$ \cite{Daou,Moll,Nandi}, Dirac semimetal PtSn$_4$ \cite{Mun2012,Gooth-PtSn4}, type-II Weyl semimetal tungsten phosphide WP$_2$ \cite{Gooth-WP2,Jaoui-WP2}, and antimony Sb \cite{Jaoui-Sb}, provided substantial evidence for viscosity-dominated electronic response. Recently, direct imaging techniques, employing scanning gate microscopy \cite{Jura,Ensslin}, a nanotube single-electron transistor \cite{Ilani}, and quantum spin magnetometry realized with nitrogen vacancy centers in diamond \cite{Yacoby,Jayich}, revealed signatures of the Poiseuille profile of electron flow in narrow graphene channels and mesoscopic Ga(Al)As. While experimental findings of electronic hydrodynamics in solid-state systems are mounting, the conceptual questions remain in particular in regard to different microscopic scattering mechanisms that govern transition to the hydrodynamic regime in different materials, and how they manifest in transport coefficients. 

\begin{figure}[h!]
  \centering
  \includegraphics[width=4.5in]{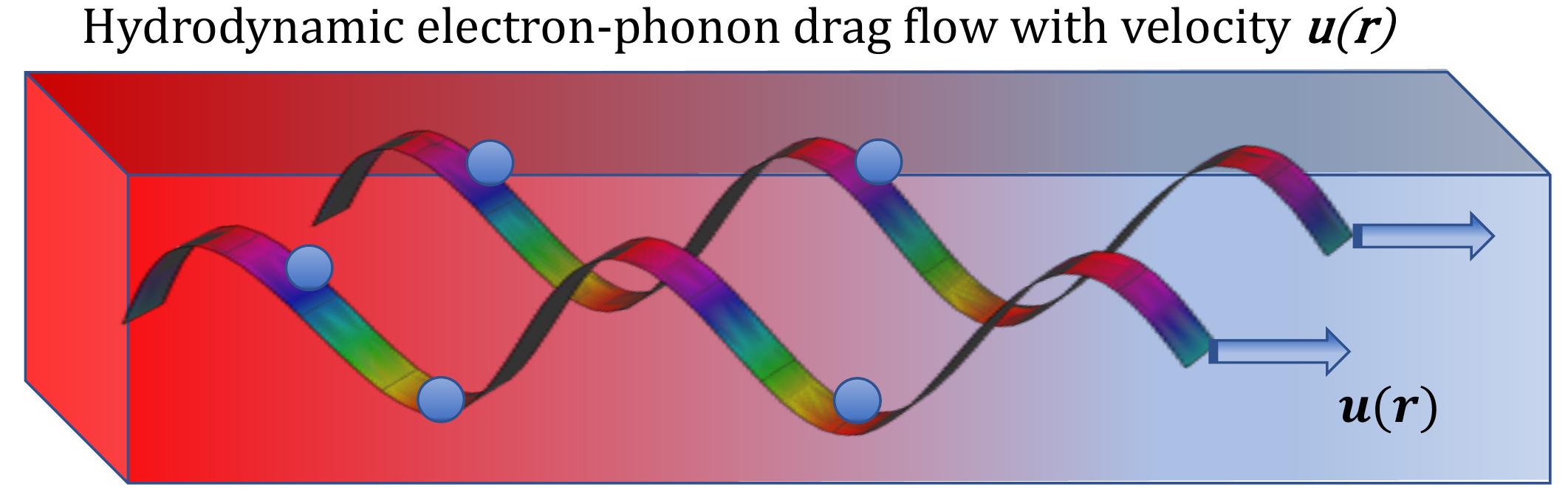}
  \includegraphics[width=1.55in]{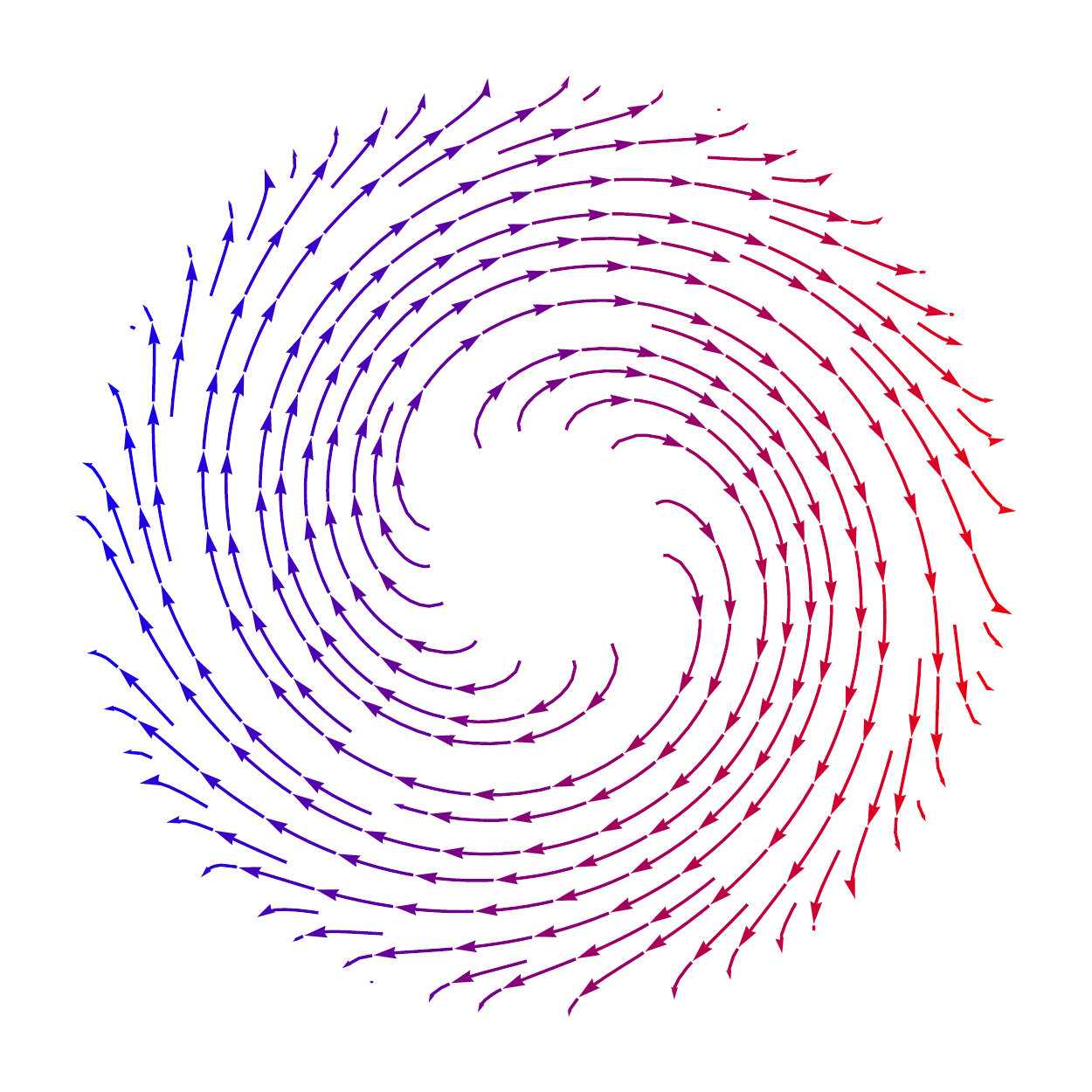}
 \caption{[Left panel]: When the total momentum of an interacting electron-phonon system  decays slowly, a coupled out-of-equilibrium state -- the electron-phonon fluid -- emerges. As a result, a joint flow-velocity $\bm{u}\left(\bm{r}\right)$ emerges as a collective hydrodynamic variable. Such a state displays viscous flow, super-diffusion in phase space and gives rise to shock-wave phenomena in the energy domain. In the figure we indicate the flow of the coupled electron wave and the moving ions that indicate the spread of acoustic waves. [Right panel]: Swirling magneto-flow profile of  $\bm{u}\left(\bm{r}\right)$  for a Corbino disc  in an applied magnetic field, discussed in Section \ref{subsec_Corbino}.}  
  \label{Fig-cover}
\end{figure}

Typically the description of electronic conduction processes in solids requires a kinetic theory that is based on the formalism of the Boltzmann equation \cite{LL-Vol10,Abrikosov,GL}. In this framework, microscopic scattering processes of momentum and energy relaxation are captured by collision terms between electrons and impurities, phonons or other relevant excitations. The electrical and thermal conductivities are then related to these microscopic length and time scales for momentum and energy relaxation. In contrast, a hydrodynamic description relies on the existence of locally conserved quantities. In this regime momentum and energy conserving electron-electron (\textit{ee}) collisions are frequent and occur on shortest length and time scales. In this picture, the resistance, for example, can be related to the electronic viscosity and the thermal conductivity \cite{AKS,Davison,Xie}. More generally, for conductors in which the underlying electron liquid lacks Galilean invariance the resistivity is determined by the entire thermoelectric matrix of the intrinsic kinetic coefficients \cite{Hartnoll,Aleiner,Lucas,PDL,LLA}. 

Since electronic scattering lengths are strongly temperature dependent and highly sensitive to the type of scattering, one often argues that the hydrodynamic regime sets in at intermediate temperatures. Indeed, at lowest temperatures when the electron-electron and electron-phonon scattering mean free paths diverge, the electronic momentum is relaxed by scattering with impurities and boundary inhomogeneities. At higher temperatures, when phonon excitation branches are activated, electron-phonon scattering is the main mechanism that relaxes both momentum and energy  of the electronic system.  In between these two limits, and provided samples of sufficient purity, there is a range of temperatures where the electron fluid attains local equilibrium on the length scale of electron-electron collisions, which is short compared to the scales at which the conservation laws break down. Then the dynamics of the electron fluid can be treated hydrodynamically. This is certainly the scenario that occurs in graphene \cite{NGMS-HydroReview,Lucas-Fong-HydroReview,Fritz2008,Mueller2009} and perhaps very high mobility semiconductor hetero-structures of moderately-strongly correlated electrons at low densities \cite{Gao,Spivak}. 

The above assertion that electron-phonon scattering is destructive for establishing electronic hydrodynamic regime by relaxing electronic momentum relies on a crucial assumption that the phonons are in thermal equilibrium. This transport situation has been considered in multiple works and much is known about momentum and energy relaxation rates from the solution of the Boltzmann equation, see for example Refs. \cite{Allen,Groeneveld,Kabanov,Sergeev}. Recently detailed \textit{ab initio} calculations provided firm results for the relevant electron-phonon scattering processes in semimetals accounting for complexities of their Fermi surfaces and microscopic details of electron-phonon coupling \cite{Coulter}. However, it was pointed out by Peierls \cite{Peierls} early on that in fact the non-equilibrium nature of the current-carrying electronic distribution should lead, through electron-phonon scattering, to a phonon distribution that is also out of equilibrium. As a consequence, the total quasi-momentum of a combined electron-phonon systems would be conserved in the absence of umklapp processes. The electrons and phonons would then drift along together, maintaining their nonzero crystal momentum and also a nonzero current, see Figure \ref{Fig-cover} for the schematic illustration. The drift velocity can be treated as an emergent hydrodynamic soft mode whose relaxation occurs at the longer time scale of umklapp scattering due to phonon nonlinearities or assisted by scattering with electrons.    

The transport theory of this phonon drag effect was developed by Gurevich in the context of thermoelectric phenomena \cite{Gurevich}. Later Gurzhi \cite{Gurzhi-UFN}, Nielsen and Shklovskii \cite{Nielsen}, and Gurevich and Shklovskii \cite{Shklovskii}, and Gurevich and Laikhtman \cite{Laikhtman} put forward hydrodynamic description of phonons in dielectrics and coupled electron-phonon liquids in metals and semiconductors (see also a detailed review \cite{Mashkevich}). Of particular relevance to our work,  Steinberg \cite{Steinberg}  and  Gurzhi and Kopeliovich \cite{Kopeliovich} considered the problem of electric conductivity of pure metals with an account of phonon drag. The electron viscosity was determined in Ref. \cite{Steinberg}, while Ref. \cite{Kopeliovich} analyzed the case of a metal with open Fermi surface consisting of large electron (or hole) groups interconnected arbitrarily by a narrow necks. In this situation the dominant cause of the low-temperature resistance is due to umklapp events occurring in collisions between electrons and phonons which remains effective down to lowest temperatures. Because of the kinematic constraints of momentum and energy conservations in scattering, the change of electron momentum in each act of collision is small and scattering occurs preferentially at small angles. As a result, electrons effectively diffuse in momentum space. This enables one to reduce the full kinetic equation to a form of a Fokker-Planck type and account for umklapp processes by imposing periodic boundary conditions on the non-equilibrium distribution function. 

In this work, we in large parts develop alternative derivations of the classic works \cite{Steinberg,Kopeliovich}, which allow us to make extensions or draw additional conclusions for the behavior of electron-phonon fluids. For example, we consider a complimentary scenario of a Peierls mechanism of umklapp scattering mediated by phonon-phonon collisions. The rate of these processes is exponential in temperature whereas the rate of normal electron-phonon collisions is a power-law. The interplay between the two leads to a pronounced peak in the temperature dependent thermopower that can be observed as one lowers the temperature. This feature is considered as one of the hallmarks of strong electron-phonon interactions as recently seen in semimetals \cite{Mun2012,Gooth-PtSn4}. Even though we face similar technical aspects of the problem as was already considered in Refs. \cite{Steinberg,Kopeliovich}, we perform a somewhat different route to analyze the problem. We do not expand the collision integrals in the limit of small momenta but rather choose to work directly with the fully coupled collision integrals. While it will not be  possible to solve these equations analytically, it is nevertheless possible to extract the main qualitative predictions from them in particular with regard to the temperature dependence of the drag viscosity and thermal conductivity of the coupled electron-phonon fluid. Furthermore, we believe that our approach may be more suitable if one wants to quantitatively describe realistic materials with a more complex shape of the Fermi surface. The rich physics that emerges if one includes such effects and anisotropies of the underlying crystal was recently elucidated in Refs. \cite{Link2018,Cook2019}.

Candidate materials for electron-phonon fluid behavior are clearly the delafossite metals PdCoO$_2$ and PtCoO$_2$ \cite{Daou,Moll,Nandi}. The temperature dependence of the bulk resistivity observed in Ref. \cite{Hicks2012} is fully consistent with phonon-drag behavior, i.e. inelastic scattering at low $T$ has an exponential temperature dependence, rather the Bloch-Gr\"uneisen behavior $\rho \propto T^5$ that occurs without phonon drag. In addition, hydrodynamic flow always requires strong momentum-conserving collisions. If collisions in the delafossite metals would be due to electron-electron scattering, their large Fermi surface would immediately give rise to equally strong umklapp processes. Hence, it seems that these systems have weak electron-electron scattering but are governed by electron-phonon scattering with phonon drag. In addition, evidence for electron-phonon fluid behavior was reported for the semimetal PtSn$_4$ \cite{Gooth-PtSn4}, another material that displays very low resistivity at low temperatures and shows a pronounced phonon drag peak in the low temperature thermopower \cite{Mun2012}. 

The remainder of the paper is organized as follows. In Sec. \eqref{Sec:KinEqs} we formulate the generic kinetic theory of coupled integro-differential equations for non-equilibrium distributions of electrons and phonons. We linearize these equations and study parity properties of the collision kernel. We also estimate rates of electron-phonon and phonon-electron collisions. Even though they originate from the same matrix elements, the respective mean free paths are parametrically different due to distinct phase space restrictions for  fermions and bosons. As a methodological exercise, we illustrate how Bloch's law for the electron-phonon resistivity follows from the solution of the integral Boltzmann equation when phonons are taken at equilibrium, and discuss how this solution is violated when complete dragging of phonons is imposed. 
Analyzing the conservation laws of the problem, we demonstrate how a joint drift velocity of the coupled electron-phonon system emerges as hydrodynamic variable, even though both constituents of the fluid have vastly different quasiparticle velocities. We finally consider a partially equilibrated case of phonon drag with rare momentum relaxing collisions and derive the hydrodynamic equation of motion for the flow of the coupled electron-phonon liquid. This analysis reveals the intrinsic viscosity and thermal conductivity in the drag regime. In Sec. \eqref{Sec:Applications} we apply this hydrodynamic description to several practical examples of viscous resistive effects and the Gurzhi effect in particular \cite{Gurzhi}. We consider flows in different geometries of a Hall bar, a quantum wire, a Corbino disk, and allow for boundary conditions with arbitrary slip length that enables us to cover the crossover from no-slip to no-stress regimes. We also consider effects of a magnetic field, and the Hall viscosity, in particular for the viscous magnetoresistance and study finite-frequency responses in the context of the skin effect. Lastly we briefly touch upon the non-equilibrium thermometry of electron-phonon collisions via shot noise in the diffusive regime. We summarize our findings in Sec. \eqref{Sec:Summary} and discuss open questions and directions for future research. Various technical calculations are delegated to several supplementary appendices that expand on properties and methods of analysis of the electron-phonon collision operator.         


\section{From kinetic to hydrodynamic theory}\label{Sec:KinEqs}

\subsection{Electron-phonon interaction}

In many practical situations and for a broad range of temperatures, the electron-phonon interaction is dominated by processes with single-phonon emissions or absorptions \cite{LL-Vol10}. Two-phonon processes could become important when one-phonon processes are forbidden or suppressed by the conservation laws or by symmetry restrictions for the transition matrix elements \cite{GL}.  
We restrict our attention to single-phonon processes exclusively. Furthermore, we will treat only the situation of scattering by long-wavelength acoustic phonons with a single electronic band. 

For spatially inhomogeneous and time-dependent conditions the coupled kinetic equations for non-equilibrium electron $n(\bm{p},\bm{r},t)$ and phonon $N(\bm{q},\bm{r},t)$ distribution functions read (hereafter $\hbar=k_B=1$): 
\begin{align}
&\frac{\partial n}{\partial t}+\bm{v}\frac{\partial n}{\partial\bm{r}}+e\bm{vE}\frac{\partial f}{\partial\varepsilon}=\St_{\mathrm{ep}}\{n,N\}+\St_{\mathrm{ei}}\{n\}, \label{KinEq-n}\\
&\frac{\partial N}{\partial t}+\bm{s}\frac{\partial N}{\partial\bm{r}}=\St_{\mathrm{pe}}\{n,N\}+\St^N_{\mathrm{pp}}\{N\}+\St^U_{\mathrm{pp}}\{N\}.\label{KinEq-N}
\end{align}  
Here $\bm{v}=\partial_{\bm{p}}\varepsilon$ and $\bm{s}=\partial_{\bm{q}}\omega$ are electron and phonon group velocities, and $\bm{E}$ is an external electric field. A finite magnetic field or temperature gradient will be added later in the text when we consider applications where this becomes necessary. In the steady-state regime the explicit time derivatives on the left hand sides vanish. In equilibrium, the fermionic and bosonic distributions are the usual Fermi-Dirac and Bose-Einstein functions   
\begin{equation}
f_\varepsilon=[\exp[(\varepsilon_{\bm{p}}-\varepsilon_F)/T]+1]^{-1}, \quad b_\omega=[\exp(\omega_{\bm{q}}/T)-1]^{-1}. 
\end{equation}
The primary focus of our attention will be the electron-phonon ($\St_{\mathrm{ep}}$) and phonon-electron ($\St_{\mathrm{pe}}$) collision integrals. The other terms such as electron-impurity $\St_{\mathrm{ei}}$, and phonon-phonon collisions, both normal type via phonon nonlinearities $\St^N_{\mathrm{pp}}$ and umklapp type $\St^U_{\mathrm{pp}}$, are kept for generality but their explicit forms will not be needed.  

The electron-phonon collision integral consists of two contributions corresponding to emission and absorption of a phonon:
\begin{align}\label{St-ep}
\St_{\mathrm{ep}}\{n,N\}=\int_{\bm{q}} W(\bm{p}|\bm{p}'\bm{q})\delta(\varepsilon_{\bm{q}}-\varepsilon_{\bm{p}'}-\omega_{\bm{q}})[n_{\bm{p}'}(1-n_{\bm{p}})N_{\bm{q}}-n_{\bm{p}}(1-n_{\bm{p}'})(1+N_{\bm{q}})] \nonumber \\ 
+\int_{\bm{q}} W(\bm{pq}|\bm{p}') \delta(\varepsilon_{\bm{p}}+\omega_{\bm{q}}-\varepsilon_{\bm{p}'})[n_{\bm{p}'}(1-n_{\bm{p}})(1+N_{\bm{q}})-n_{\bm{p}}(1-n_{\bm{p}'})N_{\bm{q}}].
\end{align}
These two terms  take care of the out-scattering and reverse in-scattering processes. In equilibrium the difference between these processes is nullified as dictated by the detailed balance condition. The momentum conservation in the first term implies $\bm{p}=\bm{p}'+\bm{q}+\textbf{g}$, while in the second $\bm{p}+\bm{q}=\bm{p}'+\textbf{g}$ where $\textbf{g}$ is reciprocal lattice vector. The phonon-electron collision integral counts the overall difference between the number of phonons emitted by electrons with momenta $\bm{p}$, as allowed by the conservation laws, and number of phonons absorbed by electron with momenta $\bm{p}'$:
\begin{align}\label{St-pe}
\St_{\mathrm{pe}}\{n,N\}=2\int_{\bm{p}} W(\bm{p}|\bm{p}'\bm{q})\delta(\varepsilon_{\bm{q}}-\varepsilon_{\bm{p}'}-\omega_{\bm{q}}) 
\left[n_{\bm{p}}(1-n_{\bm{p}'})(1+N_{\bm{q}})-n_{\bm{p}'}(1-n_{\bm{p}})N_{\bm{q}}\right].
\end{align} 
A factor of two accounts for the electron spin in these processes, and momentum conservation is implicit and fixes the momentum $\bm{p}'$. At the level of the leading Born approximation, the probabilities of scattering for direct and reverse processes are equal to each other $W(\bm{p}|\bm{p}'\bm{q})=W(\bm{pq}|\bm{p}')$. Furthermore, for the deformation potential interaction and in the long-wavelength limit, the transition probability is linearly proportional to phonon momentum $W\propto |\bm{q}|$. In what follows, we will concentrate on low-temperature processes below the scale of Debye energy, namely $T<\omega_D$. 


\subsection{Linearized collision kernels and scattering rates}

In general, it is not possible to solve the coupled nonlinear Boltzmann equations \eqref{KinEq-n} and \eqref{KinEq-N}. An analytical analysis is often restricted to the linear-response regime 
and uses solely the linearized form of the collision terms. For this purpose we assume that the distribution functions are close to their equilibrium expressions with small corrections $n=f+\delta n$ and $N=b+\delta N$. To determine the 
collision terms in Eqs. \eqref{St-ep} and \eqref{St-pe} up to   linear order in non-equilibrium corrections, it is customary to parametrize them as follows 
\begin{equation}\delta n=f(1-f)\psi=-T\frac{\partial f}{\partial\varepsilon}\psi, \quad \delta N=b(1+b)\phi=-T\frac{\partial b}{\partial\omega}\phi.
\end{equation}
This form of $\delta n$ and $\delta N$ makes it convenient to employ the detailed balance conditions under the integral. 
In addition, the expression for the entropy production in the system becomes a symmetric quadratic form in terms of $\psi$ and $\phi$, which is very useful for the variational formulation of the Boltzmann equation.

We begin with $\St_{\mathrm{pe}}$ in Eq. \eqref{St-pe} as it is simpler in structure, but the same sequence of steps will apply to the remaining collision terms. We follow the presentation given in Ref. \cite{LL-Vol10} including the notation. In the brackets of Eq. \eqref{St-pe} that account for statistical occupations we take out the product  $(1-n_{\bm{p}})(1-n_{\bm{p}'})(1+N_{\bm{q}})$ and then perform a variation of this expression with respect to the equilibrium state, which gives 
\begin{align}
\delta\St_{\mathrm{pe}}\{n,N\}=2\int_{\bm{p}} W(\bm{p}|\bm{p}'\bm{q})\delta(\varepsilon_{\bm{q}}-\varepsilon_{\bm{p}'}-\omega_{\bm{q}})
(1-f_{\varepsilon_{\bm{p}}})(1-f_{\varepsilon_{\bm{p}'}})(1+b_{\omega_{\bm{q}}})  
\delta\left[ \frac{n_{\bm{p}}}{1-n_{\bm{p}}}-\frac{n_{\bm{p}'}}{1-n_{\bm{p}'}}\frac{N_{\bm{q}}}{1+N_{\bm{q}}}\right].
\end{align}
Next we observe that 
\begin{equation}
\delta \left(\frac{n}{1-n}\right)=\frac{\delta n}{(1-f)^2}=\frac{f}{1-f}\psi,\quad 
\delta \left(\frac{N}{1+N}\right)=\frac{\delta N}{(1+b)^2}=\frac{b}{1+b}\phi,
\end{equation}
and use well-known properties between equilibrium Fermi and Bose functions (also making use of the energy-conserving delta function):
\begin{equation}\label{f-b-relations}
f_{\varepsilon_{\bm{p}}}(1-f_{\varepsilon_{\bm{p}}-\omega_{\bm{q}}})= [f_{\varepsilon_{\bm{p}}-\omega_{\bm{q}}}-f_{\varepsilon_{\bm{p}}}] b_{\omega_{\bm{q}}}, \quad
f_{\varepsilon_{\bm{p}}-\omega_{\bm{q}}}(1-f_{\varepsilon_{\bm{p}}})=
[f_{\varepsilon_{\bm{p}}-\omega_{\bm{q}}}-f_{\varepsilon_{\bm{p}}}](1+b_{\omega_{\bm{q}}}).
\end{equation} 
As a result we find 
\begin{equation}\label{St-pe-linear}
\delta\St_{\mathrm{pe}}\{\psi,\phi\}=2\int_{\bm{p}}K_-(\bm{p},\bm{q})\left[\psi_{\bm{p}}-\psi_{\bm{p}'}-\phi_{\bm{q}}\right]
\end{equation}
with the kernel 
\begin{equation}
K_\mp(\bm{p},\bm{q})=W(\bm{p}|\bm{p}'\bm{q})b_{\omega_{\bm{q}}}(1+b_{\omega_{\bm{q}}})[f_{\varepsilon_{\bm{p}}\mp\omega_{\bm{q}}}-f_{\varepsilon_{\bm{p}}}]
\delta(\varepsilon_{\bm{q}}-\varepsilon_{\bm{p}'}\mp\omega_{\bm{q}}).
\end{equation}
In complete analogy we find for the linearized version of Eq. \eqref{St-ep} the following expression 
\begin{equation}\label{St-ep-linear}
\delta\St_{\mathrm{ep}}\{\psi,\phi\}=\int_{\bm{q}} K_-(\bm{p},\bm{q})\left[\psi_{\bm{p}'}-\psi_{\bm{p}}+\phi_{\bm{q}}\right] 
-\int_{\bm{q}} K_{+}(\bm{p},\bm{q})
\left[\psi_{\bm{p}'}-\psi_{\bm{p}}-\phi_{\bm{q}}\right].
\end{equation}
The important property of these collision kernels is that they preserve the parity $\bm{q}\rightarrow -\bm{q}$ or  $\bm{p}\rightarrow -\bm{p}$ of the distribution functions. It then follows that even and odd modes of the non-equilibrium distributions are decoupled and relax on parametrically different time scales. To see this explicitly let us estimate these rates from Eqs. \eqref{St-pe-linear} and \eqref{St-ep-linear}.  The form of the out-scattering term in each of the linearized kernels suggests introducing the following rates:
\begin{equation}
\Gamma_{\mathrm{pe}}(T)=\int_p K(\bm{p},\bm{q})\sim \lambda_{\text{ep}} \frac{s}{v_F} T,  \qquad \Gamma_{\mathrm{ep}}(T) =\int_q K(\bm{p},\bm{q})\sim \lambda_{\text{ep}}  \frac{T^3}{\omega_D^2}.
\label{Gammaeppe}
\end{equation}
We suppressed here plus/minus subscript in $K(\bm{p},\bm{q})$ as phonon absorption and emission processes have the same kinematics.  Here, $\lambda_{\text{ep}}=2 D_0 p_F/(s v_F)$ is the dimensionless electron-phonon coupling constant while $D_0$ is  a constant related to the deformation potential. In what follows we estimate the given, rather distinct,  $T$-dependencies of these two rates.

At low temperatures below the scale of the Debye temperature, $T\ll\omega_D$, we have $\omega_{\bm{q}}\sim T$ and $\varepsilon_{\bm{p}}-\varepsilon_F\sim T$, so that $f_\varepsilon\sim b_\omega\sim1$. Furthermore, the typical scale of the phonon momentum is $q\sim T/s$, which is small compared to electronic momentum $p_F$, where $s$ is the sound velocity. For this reason, the delta-function in the kernel of the collision term can be simplified $\delta(\varepsilon_{\bm{p}}\pm\omega_{\bm{q}}-\varepsilon_{\bm{p}'})\approx \frac{1}{v_Fq}\delta(\cos\theta_{\bm{pq}}\pm s/v_F)$. Since $s/v_F\ll1$ it is clear that $\theta_{\bm{pq}}\sim\pi/2$ so that the phonon propagates in a direction that is almost perpendicular to the direction of the electronic momentum. In the phonon-electron scattering rate, the momentum $d^3p$ integration is taken over the volume of a layer with thickness $\sim T/v_F$ along the Fermi surface, so that $\int_p\to\nu\int d\varepsilon d\Omega$ where the solid angle is $d\Omega=2\pi\sin\theta d\theta$ and $\nu$ is the density of states at the Fermi level ($\nu=mp_F$ for a 3D metal with spherical Fermi surface). The angular average brings a factor $1/(v_Fq)\sim s/(Tv_F)$ from the delta function. Another factor of $T$ comes from $d\varepsilon$ and another $T$ from $\omega_q$ in the scattering probability $W\sim D_0(\omega_{\bm{q}}/\omega_D)$. As a result $\Gamma_{\mathrm{pe}}\sim D_0\nu(s/v_F)(T/\omega_D)$. The electron-phonon relaxation rate is estimated in exactly the same fashion, the only difference is that the integration goes over the phase space of a phonon such that $\int_q$ gives a factor $(T/s)^3$. Combined with the factor $1/(v_Fq)$ from the delta function, and a factor $\omega_{\bm{q}}/\omega_D$ from the scattering probability, this gives  
$\Gamma_{\mathrm{ep}}\simeq D_0 T^3/(v_Fs^2\omega_D)$. The rate $\Gamma_{\mathrm{ep}}$ defines the typical relaxation scale for even modes e.g. the energy relaxation.  With the above given definition of $\lambda_{\text{ep}}$ this yields our estimates for the distinct relaxation rates of electrons and phonons given in Eq. \eqref{Gammaeppe}.

The electronic momentum is relaxed on a different scale. This is not immediately clear from the form of $\Gamma_{\mathrm{ep}}$ itself but rather dictated by kinematic considerations. Indeed, during a given scattering event, the angle between the momenta of the incoming and outgoing electron is small, $\theta_{\bm{pp}'}\sim q/p_F\sim T/\omega_D$, and the change in electron momentum is $\delta p\sim q^2/p_F\ll p_F$. Thus electrons effectively diffuse in momentum space. We can easily estimate the corresponding diffusion coefficient $B$ from the  Einstein relation $\delta p^2\sim B \tau$, where $\tau \sim\Gamma^{-1}_{\mathrm{ep}}$ is the typical time scale between two consecutive collisions. This gives for $B\propto T^5$. The corresponding mean-free time for momentum relaxation, namely the time needed to change the momentum from  to its initial value, is then $\tau^{-1}_{\mathrm{ep}}\sim B/p^2_F\sim \omega_D(T/\omega_D)^5$. We can estimate the frequency of collisions of phonons with electrons in the same manner, we only need to account for the ratio between the number of electrons and the number of phonons in the region of Fermi function smearing which is of the order $\sim(T/\varepsilon_F)(T/\omega_D)^{-3}$. This implies the collision frequency per phonon occurring with the rate $\tau^{-1}_{\mathrm{pe}}\sim \omega_D(T/\varepsilon_F)$.   Hence, phonons are short-lived compared to electrons which can,  for example, be used as a justification to integrate out the lattice degrees of freedoms as  fast intermediate excitations. Such phononic states are therefore tied to the out-of-equilibrium dynamics of the electrons.  


\subsection{Bloch law and its violation under complete drag}

As a first step in our analysis, it is useful to revisit the solution of the linearized Boltzmann equation for the case of equilibrium phonons (namely neglecting the drag effect). This computation contains all the technical elements that appear in the general calculation and is helpful methodologically. When phonons are assumed to be in  equilibrium,  we can set $\phi_{\bm{q}}$ to zero in the linearized collision integral $\delta\St\{\psi,\phi\}$ of Eq. \eqref{St-ep-linear}. Thus we are looking for a solution of the following linear integral equation 
\begin{equation}\label{BKE-BG}
e\bm{vE}\frac{\partial f}{\partial\varepsilon}=\int_{\bm{q}} K_-(\bm{p},\bm{q})\left[\psi_{\bm{p}'}-\psi_{\bm{p}}\right] 
-\int_{\bm{q}} K_{+}(\bm{p},\bm{q})\left[\psi_{\bm{p}'}-\psi_{\bm{p}}\right].
\end{equation}   
The fact that the left-hand-side is odd in momentum and that kernels preserve the parity of the function tells us that $\psi_{\bm{p}}$ must be odd as well. Since $\bm{Ev}$ contains only one (first) spherical harmonic we chose a trial solution of the form 
\begin{equation}
\psi_{\bm{p}}=\frac{e\bm{vE}\tau_D}{T}\chi(\eta_{\bm{p}}),\quad \eta_{\bm{p}}=(\varepsilon_{\bm{p}}-\varepsilon_F)/T,
\end{equation} 
where time $\tau_D$ is introduced to have correct dimensionality which happens to be the characteristic relaxation time of electron-phonon collisions at $T\sim\omega_D$.
The terms with $\psi_{\bm{p}}$ and $\psi_{\bm{p}'}$ have different angular structure because the electric field has to be projected onto the initial or final momentum respectively. To resolve this difficulty we proceed as follows. Let us choose the integration $z$-axis in momentum space to be along the initial momentum $\bm{p}$. Then in the terms $\psi_{\bm{p}'}\propto (\bm{p}'\bm{E})\chi(\eta_{\bm{p}'})$ we can rewrite $\bm{p}'\bm{E}=p'_zE_z+\bm{p}'_\perp\bm{E}_\perp$ which implies an angular decomposition 
\begin{equation}
\cos\theta_{\bm{p}'\bm{E}}=\cos\theta_{\bm{pp}'}\cos\theta_{\bm{pE}}+\sin\theta_{\bm{pp}'}\sin\theta_{\bm{pE}}\cos\varphi_{\bm{p}'\bm{E}}
\end{equation}   
where $\varphi_{\bm{p}'\bm{E}}$ is the angle between projections of $\bm{p}'$ and $\bm{E}$ on the plane perpendicular to the direction of $\bm{p}$. Note that conservation of momentum and energy fixes the relationship between the angles $\theta_{\bm{pp}'}$ and $\theta_{\bm{pq}}$. Upon integration over the angle $\varphi_{\bm{p}'\bm{E}}$ the second term vanishes since we have assumed that kernels $K_\pm$ are isotropic and $\chi(\eta_{\bm{p}})$ does not depend on the direction of momentum by construction. As a result, we accumulate an extra term $\propto\cos\theta_{\bm{pp}'}\approx (1-\theta^2_{\bm{pp}'}/2)$ in the differential scattering cross-section. This is noting else but the usual angular factor in the transport scattering time. After the angular part of the integration is done, the integration over the absolute value of momentum $\bm{q}$ can be brought to a dimensionless form. Combining contributions from both $K_-$ and $K_+$,  we arrive at
\begin{equation}\label{KinEq-Bloch}
\cosh^{-2}(\eta/2)\simeq-\vartheta^3_D\int_{\eta'} K_0(\eta,\eta')\chi(\eta')+\vartheta^3_D\vartheta_F\int_{\eta'} K_1(\eta,\eta')\chi(\eta')+\vartheta^5_D\int_{\eta'} K_2(\eta,\eta')\chi(\eta'),
\end{equation}
where $\vartheta_D=T/\omega_D$, and $\vartheta_F=T/\varepsilon_F$, while 
\begin{equation}
K_k(\eta,\eta')=(\eta-\eta')^kK(\eta,\eta')
\end{equation}
for $k=0,1,2$ and also 
\begin{equation}
K(\eta,\eta')=\frac{(\eta-\eta')^2}{(1+e^{-\eta})(1+e^{-\eta'})|e^{\eta}-e^{\eta'}|}.
\end{equation}
The semi-equality sign $\simeq$ in above equation \eqref{KinEq-Bloch} implies that we kept the main parametric and functional dependences on the right-hand-side, but we suppressed all the numerical pre-factors of the order of unity in each of the three terms. Retaining these numerical factors will be done in Appendix\ref{App:BG}. Without the last two terms in Eq. \eqref{KinEq-Bloch} this equation has no solution for $\chi(\eta)$. This is the consequence of the symmetry of the kernel and the fact that uniform solution is not orthogonal to the left hand-side which is easy to check. The solution can be then found by perturbation theory treating the last two terms as corrections. The term with $K_1$ does not contribute to the leading order, as it is odd, while the second term gives 
\begin{equation}
\chi(\eta)=c/\vartheta^{5}_D
\end{equation}     
with the constant $c$ is determined by the double integral $c^{-1}=\frac{1}{4}\iint K_2(\eta,\eta')d\eta d\eta'$. With this solution at hand we can compute the electrical current 
\begin{equation}
\bm{j}=\frac{e^2\tau_D}{T}\int_{\bm{p}}\bm{v}(\bm{vE})f_{\eta_{\bm{p}}}(1-f_{\eta_{\bm{p}}})\chi(\eta_{\bm{p}})=\sigma_B\bm{E},
\label{currentBG}
\end{equation} 
with $\sigma_B=ne^2\tau_1/m$, and $\tau^{-1}_1\sim \lambda_{\text{ep}}T^5/\omega^4_D$, where dimensionless coupling constant of the electron-phonon interaction $\lambda_{\text{ep}}$ was introduced earlier in the text below Eq. \eqref{Gammaeppe}. An alternative derivation of the above formula based on the variational analysis of the functional corresponding to the Boltzmann equation \eqref{BKE-BG} is presented in Appendix \ref{App:BG}. This approach rather easily allows to fix the numerical pre-factor in $\sigma_B$ and can be naturally generalized for the calculation of other kinetic coefficients, such as thermal conductivity for example. 

As the next methodological step, it is instructive to investigate an opposite extreme limit of complete drag when the non-equilibrium electronic and bosonic distributions are locked together. 
For this case we need to solve two coupled equations
\begin{equation}
e\bm{vE}\frac{\partial f}{\partial\varepsilon}=\delta\St_{\mathrm{ep}}\{\psi,\phi\},\quad \delta\St_{\mathrm{pe}}\{\psi,\phi\}=0. 
\end{equation} 
From the second of these equations we can find bosonic function explicitly as an integral over the fermionic function [see Eq. \eqref{St-pe-linear}]
\begin{equation}
\phi_{\bm{q}}=\frac{1}{\Gamma_{\mathrm{pe}}}
\int_{\bm{p}}K_-(\bm{p},\bm{q})\left[\psi_{\bm{p}}-\psi_{\bm{p}'}\right]
\end{equation}
and insert it back into the first equation. Then repeating all the same steps as above we obtain instead of Eq. \eqref{KinEq-Bloch}
\begin{equation}\label{KinEq-Drag}
\cosh^{-2}(\eta/2)\simeq-\vartheta^3_D\int_{\eta'} K_0(\eta,\eta')\chi(\eta')+\vartheta^5_D\int_{\eta'} [K_2(\eta,\eta')-K_d(\eta,\eta')]\chi(\eta'),
\end{equation}
where a contribution with $\vartheta_F$ was omitted for brevity as it only gives a sub-leading corrections.   
The crucial new piece is the drag kernel which has the following form 
\begin{equation}
K_d=\frac{1}{2(e^\eta+1)}\int_\zeta|\zeta|^3\frac{e^\zeta+1}{e^\zeta+e^{-\eta}}\frac{e^{\eta'}}{(e^{\eta'}+e^{-\zeta})(e^{\eta'}+e^{\zeta})}.
\end{equation}
It can be shown that $\int[K_2-K_d]=0$ where the integration could be either over $\eta$ or $\eta'$. We can now integrate both sides of Eq. \eqref{KinEq-Drag} over $\eta$ to demonstrate that it has no solution. 
Physically this is the regime of infinite conductivity that can only be stabilized by momentum-relaxing collisions. 


\subsection{Super-diffusive dynamics in phase space}
\label{subsec_superdiffusion}
The same technique that we used to analyze the resistivity can be applied to determine the viscosity, as we show in the subsequent section. The scattering time $\tau_1$ listed below Eq. \eqref{currentBG} is basically $\tau_{\text{ep}}$ discussed earlier.   The subscript $l=1$ was introduced to emphasize that this time corresponds to the relaxation of a particular harmonic of the distribution function. We will determine below the viscosity that is determined by a parametrically similar time scale but that corresponds to a relaxation of the different harmonic $l=2$. In Appendix \ref{App:SuperDiff} we analyze in some detail the relaxation  $\tau_l$ of arbitrary $l$ by 
performing the angular momentum expansion of the collision term.  We obtain 
\begin{equation}
\tau_{l}^{-1}=\left\{ \begin{array}{cc}
l\left(l+1\right)240\zeta\left(5\right)\lambda_{\text{ep}}T^{5}/\omega_{D}^{4} & \:\,{\rm if}\,\,T\ll\omega_{D}/l\\
24\zeta\left(3\right)\lambda_{\text{ep}}T^{3}/\omega_{D}^{2} & \,{\rm if}\,\,\omega_{D}/l\ll T\ll\omega_{D}\\
\left(1-\delta_{l,0}\right)\left(2-\delta_{l,1}\right)\lambda_{\text{ep}} T & \:\,{\rm if}\,\,T\gg\omega_{D}
\end{array}\right.
\end{equation}
where the intermediate regime only exists at large $l$. One has to be careful with the behavior  above the Debye energy as we neglected drag corrections in the source terms and collision integrals that might correct for the numerical coefficient $2\lambda_{\text{ep}}$ for $l\geq 2$.

 These results tell us that  at lowest temperatures or for small angular momentum the collision operator can be written as an angular Laplacian $\big(\frac{1}{2}\tau^{-1}_{1}\mathbf{\hat{L}}^2\big)$, which corresponds to a diffusion on the Fermi surface. However, at any finite temperature there are angular momentum modes where we get super-diffusion. The temperature dependence of the rate of super-diffusion for $T\gg\omega_{D}$ is given by the phonon scattering rate $\Gamma_{\rm ep}\propto T^3$, introduced in Eq. \eqref{Gammaeppe}.

 We remind that the term \textit{super-diffusion} is commonly used in the literature to describe the anomalous diffusion equation 
 \begin{equation}
 (\partial_t-D\left| \Delta_{\bm p}\right|^{\mu/2})n({\bm p},t)=0
 \end{equation}
  with the exponent $\mu<2$, whereas the case $\mu>2$ is typically termed \textit{sub-diffusion}. The fractional derivative should be understood via the action of $\left| \bm{p}\right|^{\mu/2}$ in Fourier space. In our case we have $\mu=2$ at lowest temperatures while highest angular momentum states ultimately behave as  $\mu \rightarrow 0$. Notice, here diffusion takes place in phase space as a consequence of collisions. Such behavior is of importance if one analyzes the relaxation of focussed electron beams or the time dependence of heat pulses \cite{GKR,GKK,Ledwith,Kiselev2}. Related behavior was previously discussed in the context of two-dimensional Fermi liquids \cite{GKR,GKK,Ledwith} and for graphene at the neutrality point \cite{Kiselev2}. With electron-phonon fluids  we have identified three-dimensional systems that should display superdiffusive dynamics in phase space. 
  

\subsection{Emergent drift velocity and conservation laws}

In this section we use the conservation laws of the system without umklapp and impurity scattering to establish that a joint drift velocity
emerges as hydrodynamic variable. The reason for the joint drift velocity is rather transparent.  Only the total momentum $\bm{P}_{\rm tot}$ is conserved, which gives rise to only one canonically conjugate hydrodynamic variable, the drift velocity $\boldsymbol{u}\left(\boldsymbol{r}\right)$.

We start from the second law of thermodynamics as it enters the Boltzmann theory in the context of the $H$-theorem. To this end, we consider
the entropy per degree of freedom expressed in terms of the distribution functions:
\begin{equation}
s_{\bm{p}}^{{\rm el}} = - \left[n_{\boldsymbol{p}}\ln n_{\boldsymbol{p}}+\left(1-n_{\boldsymbol{p}}\right)\ln\left(1-n_{\boldsymbol{p}}\right)\right], \quad
s_{\bm{q}}^{\text{ph}}  =  -\left[N_{\bm{q}}\ln N_{\bm{q}}-\left(1+N_{\bm{q}}\right)\ln\left(1+N_{\bm{q}}\right)\right].
\end{equation}
This allows to determine the total entropy production 
\begin{equation}
Q \equiv \frac{\partial S}{\partial t}=\frac{\partial}{\partial t}\left(\int_{\boldsymbol{p}}s_{\boldsymbol{p}}^{{\rm el}}+\int_{\boldsymbol{q}}s_{\boldsymbol{q}}^{{\rm ph}}\right)=\int_{\boldsymbol{p}}\ln\left(\frac{1}{n_{\boldsymbol{p}}}-1\right)\frac{\partial n_{\boldsymbol{p}}}{\partial t}+\int_{\boldsymbol{q}}\ln\left(\frac{1}{N_{\boldsymbol{q}}}+1\right)\frac{\partial N_{\boldsymbol{q}}}{\partial t}.
\end{equation}
We can use the Boltzmann equations,  Eqs. (\ref{KinEq-n}) and (\ref{KinEq-N}),  to
express $\frac{\partial n_{\boldsymbol{p}}}{\partial t}$ and $\frac{\partial N_{\boldsymbol{q}}}{\partial t}$.
For closed systems it follows after a few steps that 
\begin{equation}
Q = -\int_{\boldsymbol{p}}\ln\left(\frac{1}{n_{\boldsymbol{p}}}-1\right)\left({\rm St}_{{\rm ep}}\left\{ n,N\right\} +{\rm St}_{{\rm ei}}\left\{ n\right\} \right)-\int_{\boldsymbol{q}}\ln\left(\frac{1}{N_{\boldsymbol{q}}}+1\right)\left({\rm St}_{{\rm pe}}\left\{ n,N\right\} +{\rm St}_{{\rm pp}}\left\{ N\right\} \right)\geq0,
 \label{QQ}
\end{equation}
where the last inequality reflects the fact that the entropy of the system
cannot decrease.  In addition we used ${\rm St}_{{\rm pp}}\left\{ N\right\} ={\rm St}_{{\rm pp}}^{N}\left\{ N\right\} +{\rm St}_{{\rm pp}}^{U}\left\{ N\right\} $
which combines normal and umklapp phonon-phonon processes.

Next we summarize the well-known implications of conservation laws.
For charge conservation we sum Eq. (\ref{KinEq-n}) over $\boldsymbol{p}$ and
obtain the continuity equation 
\begin{equation}
\frac{\partial\rho}{\partial t}+\mathbf{\nabla_{\mathbf{\boldsymbol{r}}}\cdot}\boldsymbol{j}=0
\end{equation}
since $\int_{\boldsymbol{p}}\left({\rm St}_{{\rm ep}}\left\{ n,N\right\} +{\rm St}_{{\rm ei}}\left\{ n\right\} \right)=0.$
Here we have the charge density $\rho\left(\mathbf{r},t\right)=e\int_{\mathbf{\boldsymbol{p}}}n_{\boldsymbol{p}}\left(\mathbf{r},t\right)$
and the current density $\boldsymbol{j}\left(\boldsymbol{r},t\right)=e\int_{\boldsymbol{p}}\boldsymbol{v}_{\boldsymbol{p}}n_{\boldsymbol{p}}\left(\boldsymbol{p},t\right)$.
To analyze energy conservation we introduce the energy density and
energy current of the combined system:
\begin{align}
&\varepsilon\left(\boldsymbol{r},t\right) =  2\int_{\boldsymbol{p}}\varepsilon_{\boldsymbol{p}}n_{\boldsymbol{p}}\left(\boldsymbol{r},t\right)+\int_{\boldsymbol{q}}\omega_{\boldsymbol{q}}N_{\boldsymbol{q}}\left(\boldsymbol{r},t\right), \\
&\boldsymbol{j}_{\varepsilon}\left(\boldsymbol{r},t\right)  =  2\int_{\boldsymbol{p}}\boldsymbol{v}_{\boldsymbol{p}}\varepsilon_{\boldsymbol{p}}n_{\boldsymbol{p}}\left(\boldsymbol{r},t\right)+\int_{\boldsymbol{q}}\boldsymbol{s}_{\boldsymbol{q}}\omega_{\boldsymbol{q}}N_{\boldsymbol{q}}\left(\boldsymbol{r},t\right).
\end{align}
Multiplying the Boltzmann equations by the electron and phonon energies
and integrating over momenta, we obtain 
\begin{equation}
\frac{\partial\rho_{\varepsilon}}{\partial t}+\mathbf{\nabla_{\boldsymbol{r}}\cdot}j_{\mathbf{\varepsilon}}=2 \int_{\boldsymbol{p}}(\dot{\boldsymbol{p}}\cdot\boldsymbol{v}_{\boldsymbol{p}})n_{\boldsymbol{p}}+\int_{\boldsymbol{q}}(\dot{\boldsymbol{q}}\cdot\boldsymbol{s}_{\boldsymbol{q}})N_{\boldsymbol{q}}.
\end{equation}
If there is no work done by or at the system ($\dot{\boldsymbol{p}}\cdot\boldsymbol{v}_{\boldsymbol{p}}=\dot{\boldsymbol{q}}\cdot\boldsymbol{s}_{\boldsymbol{q}}=0$)
this corresponds to the the continuity equation for the energy. It
is a consequence of the fact that the sum of $\int_{\boldsymbol{p}}\varepsilon_{\boldsymbol{p}}\left({\rm St}_{{\rm ep}}\left\{ n,N\right\} +{\rm St}_{{\rm ei}}\left\{ n\right\} \right)$
and $\int_{\mathbf{\boldsymbol{q}}}\omega_{\boldsymbol{q}}\left({\rm St}_{{\rm pe}}\left\{ n,N\right\} +{\rm St}_{{\rm pp}}\left\{ N\right\} \right)$
vanishes. Finally we consider the momentum density and momentum current:
\begin{align}
&\boldsymbol{g}\left(\boldsymbol{r},t\right)  = 2 \int_{\boldsymbol{p}}\boldsymbol{p}n_{\boldsymbol{p}}\left(\boldsymbol{r},t\right)+\int_{\boldsymbol{q}}\boldsymbol{q}N_{\boldsymbol{q}}\left(\mathbf{r},t\right),\nonumber \\
&T_{\alpha\beta}\left(\boldsymbol{r},t\right)  = 2 \int_{\boldsymbol{p}}p_{\alpha}v_{\beta}n_{\boldsymbol{p}}\left(\boldsymbol{r},t\right)+\int_{\boldsymbol{q}}q_{\alpha}s_{\beta}N_{\boldsymbol{q}}\left(\boldsymbol{r},t\right).
\end{align}
In the absence of impurity and umklapp scattering, i.e. for ${\rm St}_{{\rm ei}}\left\{ n\right\} ={\rm St}_{{\rm ep}}^{U}\left\{ n,N\right\} ={\rm St}_{{\rm pp}}^{U}\left\{ N\right\} =0$.
We obtain
\begin{equation}
\frac{\partial g_{\alpha}}{\partial t}+\frac{\partial T_{\alpha\beta}}{\partial x_{\beta}}=2\int_{\boldsymbol{p}}\dot{p}_{\alpha}n_{\boldsymbol{p}}+\int_{\boldsymbol{q}}\dot{q}_{\alpha}N_{\boldsymbol{q}},
\end{equation}
which becomes the momentum continuity equation in the absence of external
forces ($\dot{\boldsymbol{p}}=\dot{\boldsymbol{q}}=0$). The continuity
equation follows because the sum of $\int_{\boldsymbol{p}}\boldsymbol{p}\left({\rm St}_{{\rm ep}}^{N}\left\{ n,N\right\} \right)$
and $\int_{\boldsymbol{q}}\boldsymbol{q}\left({\rm St}_{{\rm pe}}^{N}\left\{ n,N\right\} +{\rm St}_{{\rm pp}}^{N}\left\{ N\right\} \right)$
vanishes. 

Let us now search for distribution functions that yield a constant entropy. Under the given
conservation laws the entropy production $Q$, as given in Eq. \eqref{QQ},  vanishes  for the distributions 
\begin{align}
&\ln\left(\frac{1}{n_{\boldsymbol{p}}}-1\right) = -\beta\left(\boldsymbol{r}\right)\mu\left(\boldsymbol{r}\right)+\beta\left(\boldsymbol{r}\right)\varepsilon_{\boldsymbol{p}}-\beta\left(\boldsymbol{r}\right)\boldsymbol{u}\left(\boldsymbol{r}\right)\cdot\boldsymbol{p},\nonumber \\
&\ln\left(\frac{1}{N_{\boldsymbol{q}}}+1\right) = \beta\left(\boldsymbol{r}\right)\omega_{\boldsymbol{q}}-\beta\left(\boldsymbol{r}\right)\boldsymbol{u}\left(\boldsymbol{r}\right)\cdot\boldsymbol{q},
\end{align}
with same $\beta\left(\boldsymbol{r}\right)$ an $\boldsymbol{u}\left(\boldsymbol{r}\right)$
in the two equations. This gives rise to local equilibrium with Fermi-Dirac
distribution function for the electrons 
\begin{eqnarray}
n_{\bm{p}}\left(\boldsymbol{r}\right) & = & \frac{1}{e^{\beta\left(\boldsymbol{r}\right)\left(\varepsilon_{\boldsymbol{p}}-\mu\left(\boldsymbol{r}\right)-\boldsymbol{u}\left(\boldsymbol{r}\right)\cdot\boldsymbol{p}\right)}+1},
\end{eqnarray}
and Bose-Einstein distribution for the phonons 
\begin{equation}
N_{\boldsymbol{q}}\left(\boldsymbol{r}\right)=\frac{1}{e^{\beta(\bm{r})\left(\omega_{\boldsymbol{q}}-\boldsymbol{u}\left(\boldsymbol{r}\right)\cdot\boldsymbol{p}\right)}-1}.
\end{equation}
Obviously we have the usual interpretation of $\beta$$\left(\boldsymbol{r}\right)$,
$\mu\left(\boldsymbol{r}\right)$ and $\boldsymbol{u}\left(\boldsymbol{r}\right)$
as local inverse temperature, chemical potential of the electrons,
and flow velocity, respectively. 

Just like the conservation of the total energy gives rise to a joint
temperature of the electrons and phonons, does  the conservation of the
total momentum yield  a joint drift velocity $\boldsymbol{u}\left(\boldsymbol{r}\right)$.
While the local equilibrium is only a solution of the Boltzmann equation
in the limit where the collision terms dominate, they do represent
a natural starting point in the limit of small Knudsen number -- i.e.
the ratio of the momentum conserving mean free path and the typical
length scale of applied forces or geometric confinement -- as employed
by the Chapman-Enskog method \cite{Cercignani}.

The hydrodynamic flow is protected by the conservation of the total momentum 
\begin{equation}
\bm{P}_{\rm tot}=\bm{P}_{\rm el}+\bm{P}_{\rm ph}
\end{equation}
and must be understood as a combined electron-phonon fluid. However, in Appendix \ref{App:Visc} we demonstrate that the primary
mechanism by which the flow gradient couples to the electron-phonon fluid is by directly affecting its electron
component. In addition we show that while phonon drag is crucial to give the viscosity a true hydrodynamic interpretation, perhaps counterintuitively, it is not important for the actual value of the viscosity. Finally, because of the larger value of the Fermi velocity and because of the different phase space nature of degenerate electrons and acoustic phonons, it holds that the momentum current is also dominated by the electronic system.


\subsection{Hydrodynamic electron-phonon drag viscosity}

Provided that momentum-conserving electron-phonon collisions are the most frequent, the regime of phonon drag can be characterized by an emergent hydrodynamic mode, which is the drift velocity of electrons and phonons. Indeed, both collision terms $\St_{\mathrm{ep}}\{n,N\}$ and $\St_{\mathrm{pe}}\{n,N\}$ are simultaneously solved by a distribution function with the finite boost $n(\bm{p},\bm{r})=f(\varepsilon_{\bm{p}}-\bm{pu}(\bm{r}))$ and  $N(\bm{p},\bm{r})=b(\omega_{\bm{q}}-\bm{qu}(\bm{r}))$. In the previous sub-section we discuss the origin of the joint drift velocity as conjugated variable to the conserved total momentum in some detail.

To determine the equation of motion for $\bm{u}(\bm{r})$ we follow the approach of Gurzhi \cite{Gurzhi-UFN} who solved the kinetic equations by the method of consecutive approximations. The accuracy of the method is controlled by the ratio between momentum-conserving and momentum-relaxing scattering lengths. We seek  the non-equilibrium distribution functions in the form of a formal series expansion: $n=f+\delta n_1+\delta n_2+\ldots ... $ and $N=b+\delta N_1+\delta N_2$.   
To the first order we obtain two equations:
\begin{equation}
\bm{v}\frac{\partial f}{\partial\bm{r}}=\delta\St_{\mathrm{ep}}\{\delta n_1,\delta N_1\},\quad \bm{s}\frac{\partial b}{\partial\bm{r}}=\delta\St_{\mathrm{pe}}\{\delta n_1,\delta N_1\},
\end{equation}
where linearized collision kernels are given by Eqs.  \eqref{St-pe-linear} and \eqref{St-ep-linear}. The contribution from the collision term with normal phonon processes, governed by $\delta\St^N\{\delta N_1\}$, can be neglected as it has a subdominant temperature dependence in comparison with phonon-electron collisions. Since the spatial dependency of the distribution is contained in the velocity field $\bm{u}(\bm{r})$ we search for a solution of the form 
\begin{align}\label{n-1}
\delta n_1=-v_ip_j\frac{\partial u_j}{\partial r_i}\frac{\partial f}{\partial\varepsilon}\tau_D(\omega_D/T)^3\chi(\eta_{\bm{p}}),\\
\delta N_1=-s_iq_j\frac{\partial u_j}{\partial r_i}\tau_D(T/ms^2)\frac{\partial b}{\partial\omega}\phi(\zeta_{\bm{q}}),\label{N-1}
\end{align}
where $\zeta_{\bm{q}}=\omega_{\bm{q}}/T$. Again repeating all the same technical steps from the previous section, where we discussed Bloch's solution of the linearized kinetic equations, we find two coupled integral equations for the non-equilibrium distributions $\psi$ and $\phi$:
\begin{align}
&\cosh^{-2}(\eta/2)\simeq-\int_{\eta'}K_0(\eta,\eta')\chi(\eta')+\vartheta^2_D\int_{\eta'}K_2(\eta,\eta')\chi(\eta')-\vartheta^2_D\int_{\eta'}K_1(\eta,\eta')\phi(\eta-\eta'), \\ 
&\zeta/(e^\zeta-1)\simeq\vartheta^2_D\zeta^2\phi(\zeta)+(1-\vartheta^2_D\zeta^2)\int_{\zeta'}Q(\zeta,\zeta')\chi(\zeta'),
\end{align} 
where 
\begin{equation}
Q(\zeta,\zeta')=\frac{(e^\zeta-1)(e^{\zeta'}-1)}{(e^\zeta+1)(e^\zeta+e^{\zeta'})(e^\zeta+e^{-\zeta'})}.
\end{equation}
We were unsuccessful in finding an analytical solution of these equations. However, exploring the smallness of $\vartheta_D\ll1$ it is possible to show that $\chi(\eta)\sim \vartheta^{-2}_{D}$.  

At the second order of the expansion, the set of equations  takes  the form 
\begin{align}\label{n-2}
&\bm{v}\frac{\partial\delta n_1}{\partial\bm{r}}+e\bm{vE}\frac{\partial f}{\partial\varepsilon}=\delta\St_{\mathrm{ep}}\{\delta n_2,\delta N_2\}+\delta\St_{\mathrm{ei}}\{f\},\\
&\bm{s}\frac{\partial\delta N_1}{\partial\bm{r}}=\delta\St_{\mathrm{pe}}\{\delta n_2,\delta N_2\}+\delta\St^U_{\mathrm{pp}}\{b\}.\label{N-2}
\end{align}  
It is important to emphasize at this point that $\delta\St_{\mathrm{ei}}\{f\}\propto \bm{u}(\bm{r})$ and similarly  $\delta\St^U_{\mathrm{pp}}\{f\}\propto \bm{u}(\bm{r})$ as these two terms capture momentum-relaxing collisions and as such will define the relaxation of $\bm{u}$. Finally, we use the explicit form of $\delta n_1$ from Eq. \eqref{n-1}, multiply Eq. \eqref{n-2} by $\bm{p}$ and integrate both sides over momentum. Similarly we use $\delta N_1$ from Eq. \eqref{N-1} in Eq. \eqref{N-2}, multiply by $\bm{q}$ and integrate both sides. We then add together these equations and obtain the desired hydrodynamic equation for $\bm{u}(\bm{r})$ (see also Refs. \cite{Gurzhi-UFN,Kopeliovich}):  
\begin{equation}\label{u}
\nu\nabla^2\bm{u}+e\bm{E}/m=\bm{u}/\tau_{\mathrm{MR}}. 
\end{equation}
Here momentum-relaxation time $\tau^{-1}_{\mathrm{MR}}=\tau^{-1}_{\mathrm{ei}}+\tau^{-1}_{U}$ is given by the sum of two terms due to electron-impurity and phonon umklapp scattering. While the former is temperature independent, the latter has steep exponential behavior $\tau^{-1}_{U}\propto (T/\omega_D)^4(\tau^{U}_{\text{pp}})^{-1}$, with $(\tau^{U}_{\text{pp}})^{-1}\propto \exp(-\gamma\omega_D/T)$ and $\gamma\sim1$. The kinematic viscosity of the electron-phonon fluid $\nu=\eta_{\text{ep}}/mn$ in Eq. \eqref{u} is expressed in terms of the corresponding shear viscosity in a standard way:
\begin{equation}\label{ep-viscosity}
\eta_{\text{ep}}=\frac{1}{5}mnv^2_F\tau_2,\quad \tau^{-1}_2=1440\zeta(5)\lambda_{\text{ep}}T^5/\omega^4_D.  
\end{equation}
For the detailed derivation of Eq. \eqref{ep-viscosity} see Appendix \ref{App:Visc}. Notice that the functional form of Eq. \eqref{u} is formally identical to the equation of motion of an electron fluid where the hydrodynamic regime is established by electron-electron collisions. The difference is only in the temperature dependence of the viscosity, i.e. of the relaxation time $\tau_2$.  The electron-phonon collisions that give rise to a $T^5$ Bloch-Gr\"uneisen law in the resistivity of the kinetic regime are the same processes that determine the viscosity $\eta_{\text{ep}} \propto T^{-5}$ in the hydrodynamic regime. This parallels the electron-electron hydrodynamic regime where the $T^2$ term in the resistivity translates into $\eta_{\text{ee}} \propto T^{-2}$ for the electron viscosity \cite{AbrikosKhalat}. In closing this section we also wish to draw attention to an analogy between phonon drag viscosity and recently studied Coulomb drag viscosity contribution \cite{Galitski}, and its relation to hydrodynamic drag resistivity in the transport properties of interactively coupled double-layers \cite{Chen}. 


\subsection{Thermal conductivity and the Lorentz ratio in a drag regime}

The theory of thermal conductivity in the hydrodynamic regime of a phonon gas was put forward in pioneering works of Callaway \cite{Callaway} and Gurzhi \cite{Gurzhi-ThermalCond} (the classical review on the topic can be found in Ref. \cite{Thellung}, whereas a concise summary of the field with the modern perspective can be found in Ref. \cite{Lee-Li}). These authors carefully analyzed the interplay of various scattering processes including (i) sample boundary scattering, described by a constant relaxation time; (ii) three-phonon nonlinearities, whose relaxation time is a power-law of temperature; (iii) impurity scattering; (iv) umklapp processes with an exponential relaxation time. The resulting thermal conductivity was shown to exhibit fairly complicated non-monotonic behavior. Recently phonon-mediated heat diffusion in insulators received a renewed attention and interest triggered by a realization of apparently universal bound controlled by the Plankian time scale, $\tau_{\text{Pl}}\sim (\hbar/k_B T)$, quantum mechanical bound on sound velocity \cite{Kapitulnik-Planckian,Hartnoll-Planckian}, and generalization of Fourier's law into viscous heat equations \cite{Cepellotti}. In this section we consider the problem of thermal conduction from the perspective of mutual electron-phonon drag and reveal its distinct properties. The corresponding electron-phonon bound on thermal diffusion can be analyzed in a similar spirit as it was done recently in the context of the Coulomb drag problem \cite{Holder-DragBounds}. 

The starting point of our treatment is the same set of linearized coupled integro-differential Boltzmann equations as used in the case of conductivity and viscosity calculations in previous sections. The only difference is that we are looking now at the response to the temperature gradient $\nabla_{\bm{r}} T$, thus we have 
\begin{equation}
-\frac{\varepsilon_{\bm{p}}}{T}\frac{\partial f}{\partial\varepsilon_{\bm{p}}}(\bm{v}_{\bm{p}}\nabla_{\bm{r}}T)
=\delta\St_{\text{ep}}\{\delta n_1,\delta N_1\},\quad -\frac{\omega_{\bm{q}}}{T}\frac{\partial b}{\partial\omega_{\bm{q}}}(\bm{s}_{\bm{q}}\nabla_{\bm{r}}T)=\delta\St_{\text{pe}}\{\delta n_1,\delta N_1\}.
\end{equation}
It is clear that in the linear response analysis the non-equilibrium corrections to electron and phonon distribution functions are proportional to the thermal bias, namely $\{\delta n_1,\delta N_1\}\propto \nabla T$. Provided that a solution is found, the heat current can be computed in accordance with the usual kinetic formula   
\begin{equation}\label{j-E}
\bm{j}_\varepsilon=\int_{\bm{p}}\bm{v}_{\bm{p}}\varepsilon_{\bm{p}}\delta n_1+\int_{\bm{q}}\bm{s}_{\bm{q}}\omega_{\bm{q}}\delta N_1=-\kappa_{\text{ep}}\nabla_{\bm{r}} T, 
\end{equation}
that thus defines the electron-phonon drag thermal conductivity $\kappa_{\text{ep}}$. Just like in the case for the electron viscosity calculation, discussed in the Appendix \ref{App:Visc}, we can first solve for the non-equilibrium phonon distribution $\delta N_1$ in terms of yet unknown $\delta n_1$, and insert the result into the Boltzmann equation for the electrons. This yields then purely electronic Boltzmann
equation of the type
\begin{equation}
\bm{R}_{\bm{p}}\cdot \nabla_{\bm{r}}T=\delta\St_{\text{el}}\{\delta n_1\}. 
\end{equation}
The source term on the right-hand-side $\bm{R}_{\bm{p}}=-\bm{v}_{\bm{p}}(\varepsilon_{\bm{p}}/T)(\partial f/\partial\varepsilon_{\bm{p}})+\delta \bm{R}_{\bm{p}}$ is renormalized by the drag effect. The collision term $\delta\St\{\delta n_1\}$ also contains an additional correction. The analysis of the second term $\delta\bm{R}_{\bm{p}}$ yields the conclusions that it can be neglected at temperatures $T\ll\omega_D$. The subsequent analysis of the collision term is analogous to the one for the viscosity and yields for the thermal conductivity the result (see Appendix \ref{App:ThCond} for further details) 
\begin{equation}\label{kappa-ep}
\kappa_{\text{ep}}=\frac{1}{3}v^2_Fc_{\text{el}}(T)\tau_{E},\quad \tau^{-1}_{E}= 480\zeta(5)\lambda_{\text{ep}}T^3/\omega^2_D,\quad T\ll\omega_D,
\end{equation}
with the electronic heat capacity $c_{\text{el}}(T)$. For higher temperatures it holds that $\tau^{-1}_{E}\simeq \lambda_{\text{ep}}T$.
With the linear low-$T$ heat capacity $c_{\text{el}}\simeq \gamma_s T$, where $\gamma_s$ is the usual Sommerfeld coefficient, it follows for the thermal conductivity $\kappa_{\text{ep}}\propto 1/T^2$. This is distinct from the thermal conductivity of a Fermi liquid $\kappa_{\text{ee}}\propto 1/T$ \cite{AbrikosKhalat} and would naturally lead to a temperature dependent Lorentz ratio, $L(T)=\kappa/\sigma T$, quite distinct from the universal Sommerfeld bound of $\pi^2/3e^2$ in the Wiedemann-Franz law. We note that thermal conductivity has been measured recently in Refs. \cite{Daou,Gooth-PtSn4} in a phonon drag regime driven by normal electron-phonon scattering processes.  The scaling consistent with $T^{-2}$-behavior in the intermediate range of temperatures was indeed observed in PtSn$_4$ \cite{Gooth-PtSn4}. Additionally, hydrodynamic features due to electron viscosity accompanied by the size-dependent departure from the Wiedemann-Franz law, expected in the hydrodynamic picture, were observed in recent thermal resistivity measurements in semi-metallic antimony Sb \cite{Jaoui-Sb}. Similar thermal transport anomalies were also reported in WP$_2$ \cite{Gooth-WP2} and analyzed theoretically in Ref. \cite{Li-Maslov}.  


\begin{figure}[t!]
  \centering
  \includegraphics[width=3in]{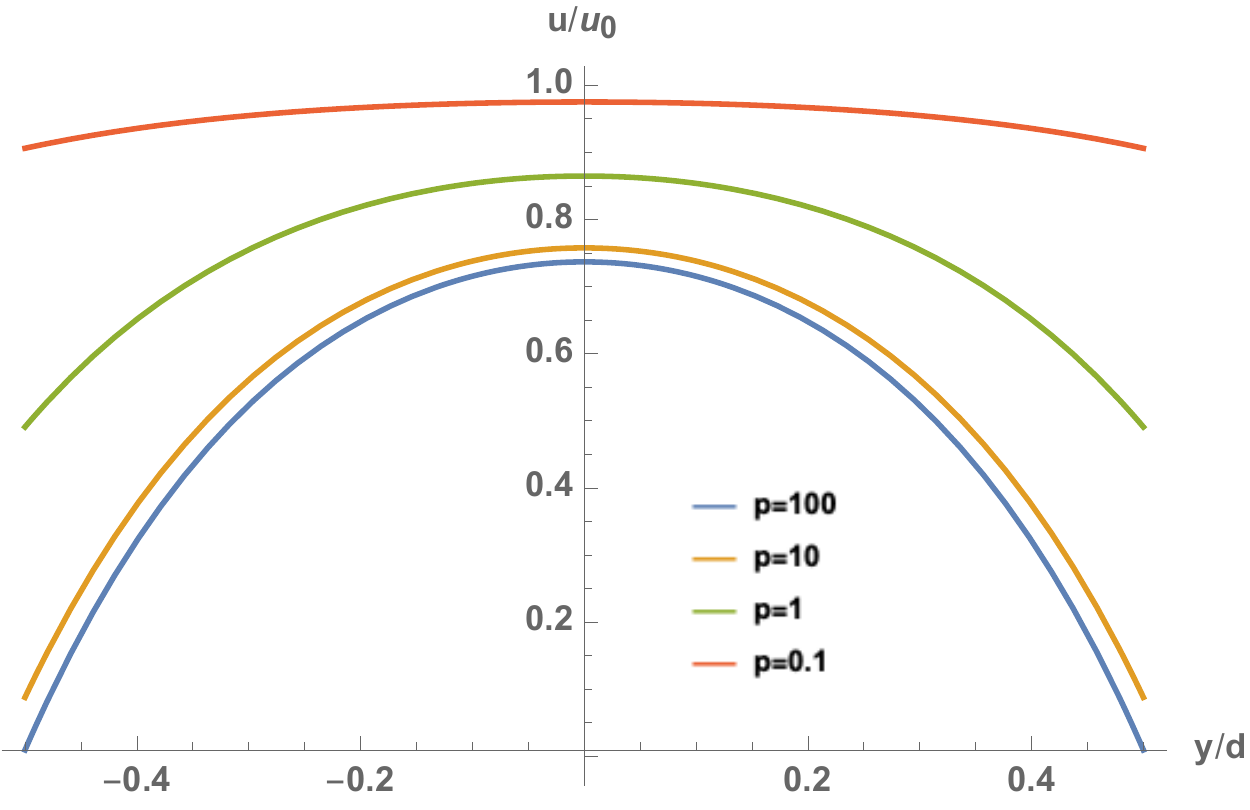}
  \includegraphics[width=3in]{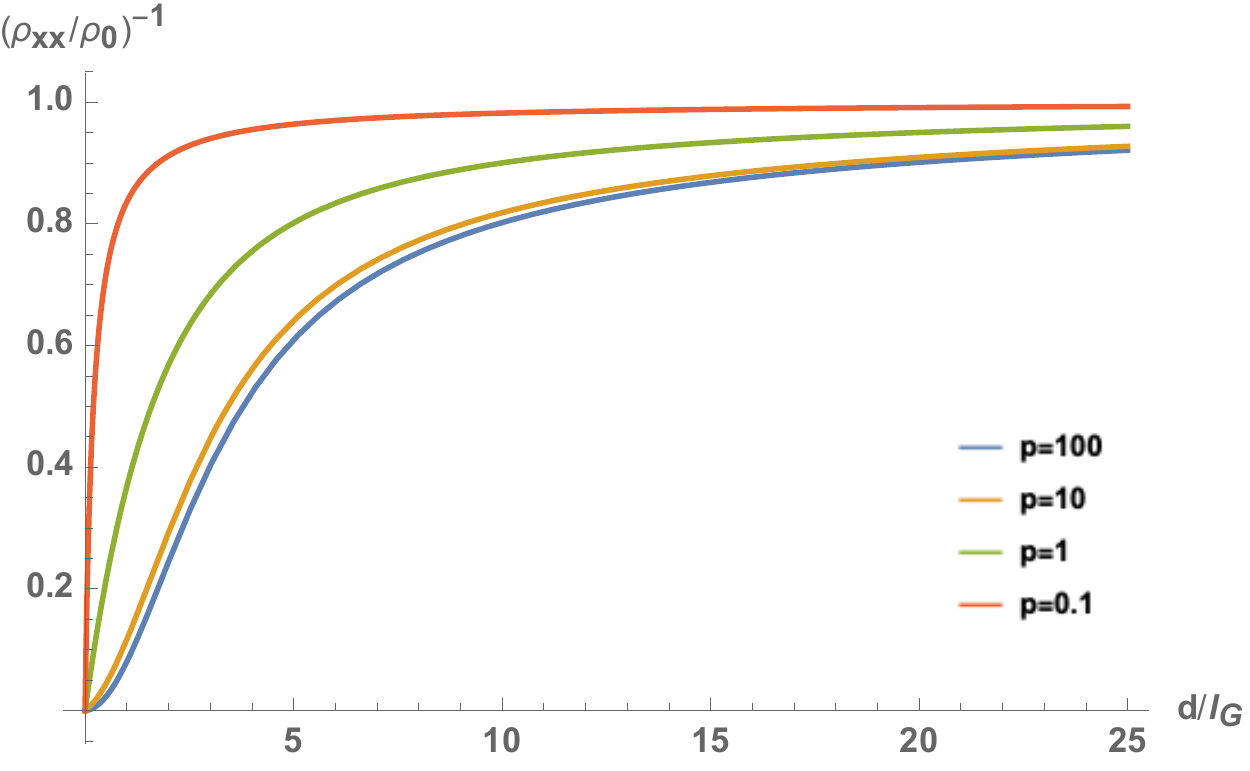}
 \caption{[Left panel]: Spatial profile of the hydrodynamic flow field in the slab geometry for $d/l_{\mathrm{G}}=4$ in the crossover regime from no-slip to no-stress boundary conditions. [Right panel]: Dependence of the resistivity as a function of the channel width normalized to the Gurzhi length plotted for several different values between Gurzhi length and slip length. }  
  \label{Fig-u}
\end{figure}

\section{Applications}\label{Sec:Applications}

\subsection{Gurzhi resistance at arbitrary slip length}

As a first application let us consider hydrodynamic flow in a  two-dimensional slab geometry of width $d$ where the flow occurs in the $x$-direction such that the velocity field $\bm{u}\left(\bm{r}\right)=(u(y),0)$ has a nontrivial profile along the  $y$-direction, where the electric field $\bm{E}=(E_x,0)$ is directed along $x$-direction. The equation of motion \eqref{u} then becomes
\begin{equation}
\nu\frac{d^2u}{dy^2}+eE_x/m=u/\tau_{\mathrm{MR}}.
\end{equation}  
This equation should be supplemented by a boundary condition. We use a generic one allowing for an arbitrary slip length $l_{\mathrm{S}}$ \cite{Kiselev,Moessner2}
\begin{equation}\label{u-bc}
\left(\frac{du}{dy}\right)_{y=\pm d/2}=\mp \frac{u(\pm d/2)}{l_{\mathrm{S}}}. 
\end{equation} 
Solving this linear differential equation we find a flow profile 
\begin{equation}
u(y)=u_0\left[1-2p\frac{(1+p)e^{w/2}+(1-p)e^{-w/2}}{(1+p)^2e^w-(1-p)^2e^{-w}}\cosh\frac{y}{l_{\mathrm{G}}}\right].
\end{equation}
Here we introduced the characteristic steady state velocity $u_0$, the Gurzhi length $l_{\mathrm{G}}$, and two dimensionless parameters $p, w$: 
\begin{equation}
u_0=\frac{eE_x\tau_{\mathrm{MR}}}{m}, \quad l_{\mathrm{G}}=\sqrt{\nu\tau_{\mathrm{MR}}},\quad p=l_{\mathrm{G}}/l_{\mathrm{S}},\quad w=d/l_{\mathrm{G}}. 
\end{equation}
The no-slip boundary condition corresponds to the limit where $p\to\infty$, whereas the opposite limit $p\to0$ defines the no-stress regime. The flow profiles at different values of $p$ are illustrated in Fig.  \ref{Fig-u}. 
We introduce the average flow velocity across the channel
\begin{equation}
\bar{u}=\frac{1}{d}\int^{d/2}_{-d/2}u(y)dy.
\end{equation}   
This expression enables us to find current density $j_x=en\bar{u}$ and consequently resistance along the channel 
\begin{equation}
\rho^{-1}_{xx}\!=\!\rho^{-1}_0\left[1-\frac{4p}{w}\frac{(1+p)e^{w/2}+(1-p)e^{-w/2}}{(1+p)^2e^w-(1-p)^2e^{-w}}\sinh\frac{w}{2}\right],
\end{equation}
where $ \rho^{-1}_0=e^2n\tau_{\mathrm{MR}}/m$ is the familiar formula of the Drude resistivity. This result simplifies in the limit of no slip $p\to\infty$ \cite{Gurzhi,Alekseev}
\begin{equation}
\rho^{-1}_{xx}=\rho^{-1}_0\left[1-\frac{2}{w}\tanh\frac{w}{2}\right].
\end{equation}
For a wide channel, $d\gg l_{\mathrm{G}}$, the resistance saturates to its bulk value $\rho_0$ which is governed by the momentum-relaxing time. 
In contrast, for a narrow channel, $d\ll l_{\mathrm{G}}$, the resistivity is determined by momentum conserving electron-phonon collisions and inversely proportionally to the channel width 
as expected for the Poiseuille flow $\rho_{xx}\simeq(p_F/e^2n)(l_{\mathrm{MC}}/d)^2$. This defines the regime of the Gurzhi effect \cite{Gurzhi} when the resistance drops with increasing temperature
as controlled by the momentum conserving length scale $l_{\mathrm{MC}}=v_F\tau_{\mathrm{ep}}$.   

As the next step, we briefly investigate the sensitivity of these results to the geometry of the conducting channel. For this purpose we look at the quantum wire (cylindrical geometry) of radius $d$. Using the Laplacian in radial coordinates the equation of motion and boundary condition take the form 
\begin{equation}
\frac{1}{r}\frac{d}{dr}\left(r\frac{du}{dr}\right)-\frac{u}{l^2_{\mathrm{G}}}=-eE_x/m\nu,  \quad
\quad (du/dr)_{r=d}=-u(d)/l_{\mathrm{S}},
\end{equation}
where $\bm{u}\left(\bm{r}\right)=(u(r),0,0)$. This equation is solved in terms of the modified Bessel functions of zero index. However, for a bounded solution at the origin we must retain only $\mathrm{I}_0$ function but not $\mathrm{K}_0$. Recalling then the property of the derivative that $\mathrm{I}'_0(z)=\mathrm{I}_1(z)$ we find 
\begin{equation}
u(r)=u_0\left[1-\frac{p\mathrm{I}_0(r/l_{\mathrm{G}})}{\mathrm{I}_1(w)+p \mathrm{I}_0(w)}\right].
\end{equation}
Averaging this expression over the wire cross-section and recalling the integral property 
\begin{equation}
\int^{z}_{0}r\mathrm{I}_0(r)dr=z \mathrm{I}_1(z)
\end{equation}
we find wire resistivity in the form
\begin{equation}
\rho^{-1}_{xx}=\rho^{-1}_0\left[1-\frac{2p}{w}\frac{\mathrm{I}_1(w)}{\mathrm{I}_1(w)+p \mathrm{I}_0(w)}\right]. 
\end{equation}
The flow profile is analogous to that of a slab presented in Fig. \eqref{Fig-u} with the only difference that it looks flatter at the center of the wire. The resistance also exhibits the same  dependency on the ratio $d/l_{\mathrm{G}}$. The only difference is numerical coefficients of the order unity that occur in the respective asymptotic limits.  

\begin{figure}
  \centering
  \includegraphics[width=3.15in]{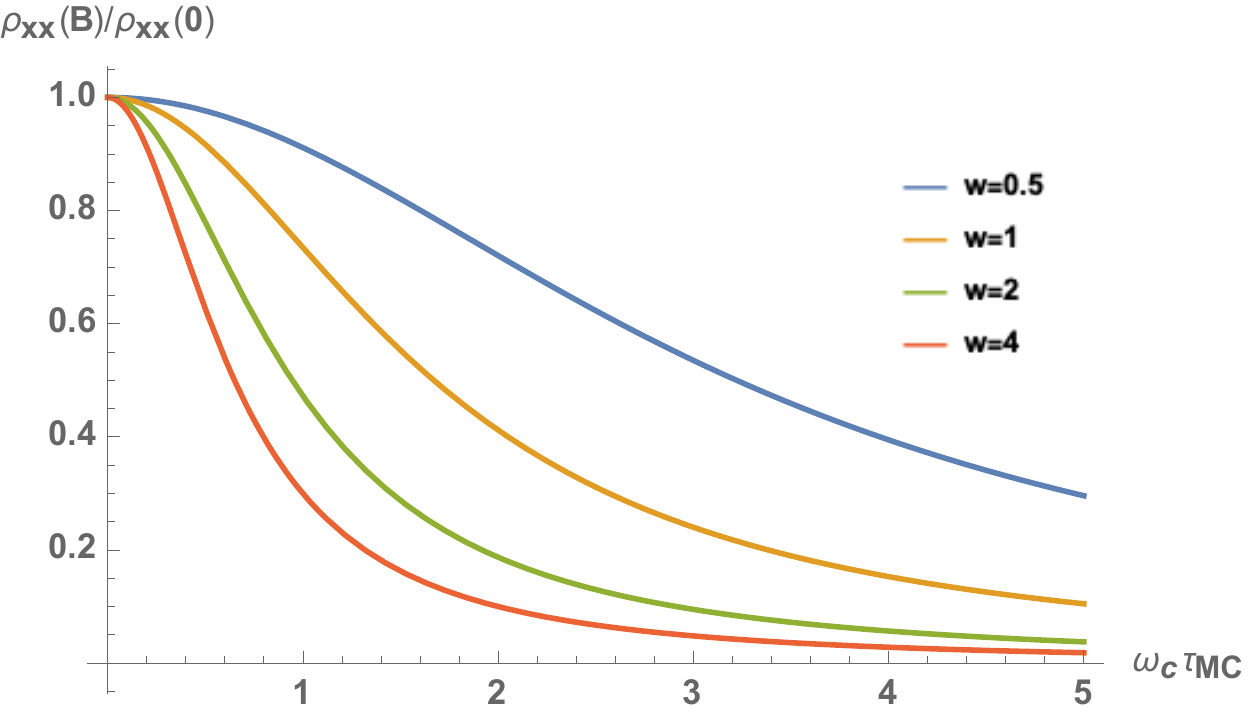}
  \includegraphics[width=3.15in]{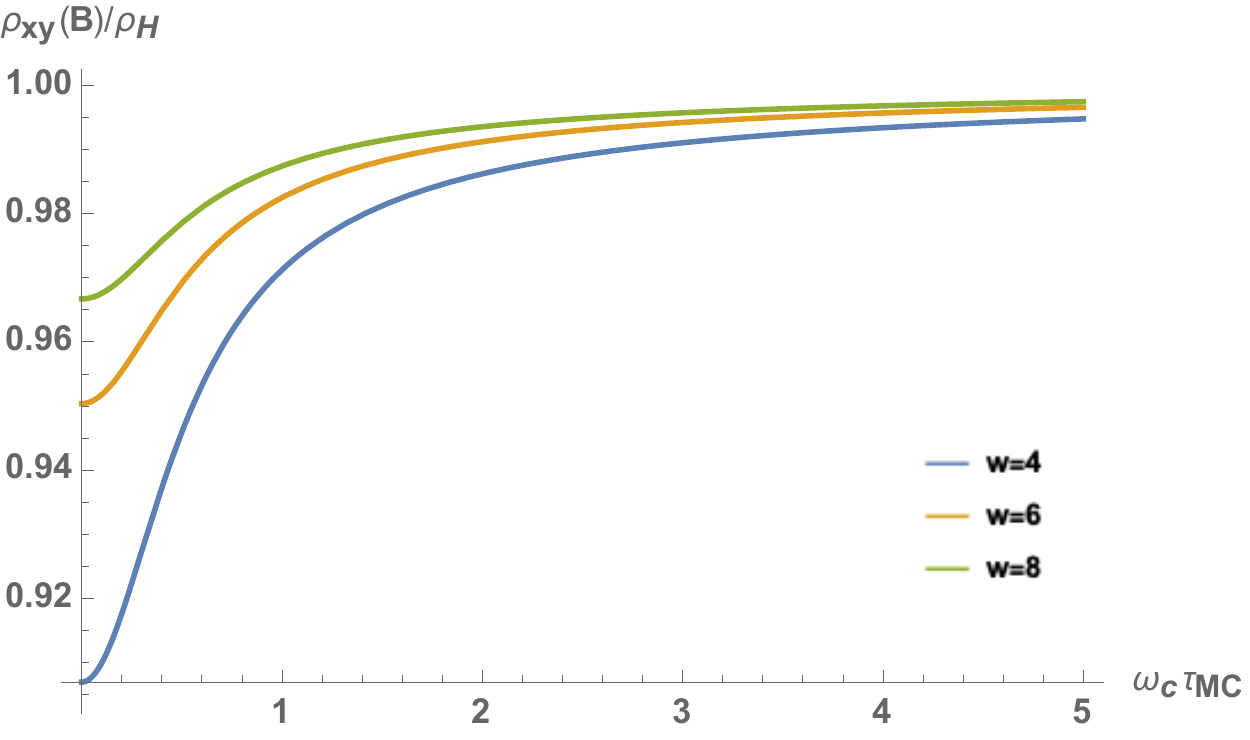}
 \caption{[Left panel]: Field dependence of the diagonal resistivity for different channel width aspect ratios. [Right panel]: Field dependence for the Hall resistivity normalized to the classical Hall resistance.}  
  \label{Fig-R-B}
\end{figure}


\subsection{Magnetoresistance and Hall resistance} 

In the presence of an external magnetic field we need to add the Lorentz force as well as a Hall viscosity $\nu_{\mathrm{H}}$ into the equation of motion. For a steady flow we thus have \cite{Alekseev,Scaffidi,Gromov,Polini,Hankiewitcz}
\begin{equation}\label{NS-MHD}
\nu\nabla^2\bm{u}+\nu_{\mathrm{H}}[\nabla^2\bm{u}\times \bm{e}_{B}]+e(\bm{E}+[\bm{u}\times\bm{B}])/m=\bm{u}/\tau_{\mathrm{MR}},
\end{equation}
where $\bm{e}_B$ is the unit vector along the magnetic field. In the semiclassical approximation, the Hall viscosity can be derived in a manner similar to the classical work of Steinberg \cite{Steinberg} (see also the recent discussions by Alekseev \cite{Alekseev} and Scaffidi {\em et al.} \cite{Scaffidi}). The difference is that for our case it is assumed that the kinematic viscosity is dominated by electron-phonon collisions instead of electron-electron collisions. 

For a Hall bar strip geometry with magnetic field along the $z$-axis there is no flow in the $y$-direction due to the build up of an electric field that compensates for the Lorentz force in the classical Hall effect.  The resulting equations of motion  read  
\begin{equation}
\nu\frac{d^2u_x}{dy^2}+eE_x/m=u_x/\tau_{\mathrm{MR}},\quad
-\nu_{\mathrm{H}}\frac{d^2u_x}{dy^2}+eE_y/m=\omega_c u_x,
\end{equation}
where $\omega_c=eB/m$ is the cyclotron frequency. The first equation is structurally unchanged as compared to the case of no field, so is solved exactly as in the previous section. To find $E_y$ in the second equation, we integrate this equation over the strip width and get 
\begin{equation}
-\frac{\nu_{\mathrm{H}}}{d}\left[\left(\frac{du_x}{dy}\right)_{y=d/2}-\left(\frac{du_x}{dy}\right)_{y=-d/2}\right]+eE_y/m=\omega_c\bar{u}_x
\end{equation}
From the boundary conditions, we can express derivatives of the velocity field in terms of the velocity itself and the slip length  
\begin{equation}
\frac{2\nu_{\mathrm{H}}}{dl_{\mathrm{S}}}u_x(d/2)+\frac{eE_y}{m}=\omega_c\bar{u}_x
\end{equation}
This equation yields the Hall field (and voltage) and thus gives us the transverse resistivity
\begin{equation}
\rho_{xy}=\rho_\mathrm{H}\left[1-\frac{\nu_{\mathrm{H}}}{d^2\omega_c}H(p,w)\right],\quad \rho_\mathrm{H}=\frac{B}{en}.
\end{equation}
The dimensionless function $H(p,w)=2p wu_x(d/2)/\bar{u}_x$ can be found from the longitudinal flow profile of the velocity field and is given by 
\begin{equation}
H(p,w)=2p w\frac{1-2p P(p,w)\cosh(w/2)}{1-(4p /w)P(p,w)\sinh(w/2)},
\end{equation}
where 
\begin{equation}
P(p,w)=\frac{(1+p)e^{w/2}+(1-p)e^{-w/2}}{(1+p)^2e^w-(1-p)^2e^{-w}}.
\end{equation}
The Hall resistance takes a particularly simple form in the no-slip limit where 
\begin{equation}
H=\frac{2w\tanh(w/2)}{1-(2/w)\tanh(w/2)}.
\end{equation}
In the weak-field limit, taking $\nu_\mathrm{H}\simeq\nu(\omega_c\tau_{\mathrm{MC}})$, where $\tau_{\mathrm{MC}}$ is the momentum conserving time scale, given by electron-phonon collisions in our case, 
we estimate the correction to the Hall resistivity to be of the form 
\begin{equation}
\frac{\delta\rho_{xy}}{\rho_{\mathrm{H}}}\simeq-\left(\frac{l_{\mathrm{MC}}}{d}\right)^2\frac{(2d/l_\mathrm{G})\tanh(d/2l_{\mathrm{G}})}{1-(2l_{\mathrm{G}}/d)\tanh(d/2l_{\mathrm{G}})}.
\end{equation}  
We remind that the underlying assumption for the length scales is such that $l_{\mathrm{MC}}\ll d\ll l_{\mathrm{MR}}$. Note that the Gurzhi length can be equivalently presented as $l_\mathrm{G}=\sqrt{l_{\mathrm{MC}}l_{\mathrm{MR}}}$ such that, in principle, the relationship between $d$ and $l_\mathrm{G}$ can be arbitrary. Having this in mind we conclude that the correction $\delta\rho_{xy}$ is universal in the narrow channel when $d\ll l_{\mathrm{G}}$ where $\delta\rho_{xy}/\rho_{\mathrm{H}}\simeq -(l_{\mathrm{MC}}/d)^2$ while it scales as $\delta\rho_{xy}/\rho_{\mathrm{H}}\simeq -l^2_{\mathrm{MC}}/dl_{\mathrm{G}}$ in the opposite limit. The field dependence of both, the diagonal and the Hall resistivities in the semiclassical limit is illustrated in Fig. \eqref{Fig-R-B} for different aspect ratios of the Hall bar channel  and different ratios of the channel width and the Gurzhi length, respectively.


\subsection{Stokes-to-Ohm crossover in a swirling magneto-flow}
\label{subsec_Corbino}

The 2D cylindrical geometry of a Corbino disk with inner radius $r_1$ and outer radius $r_2$ also attracts considerable attention. It was recently suggested that the electronic shear viscosity can be measured with this device in the response to an alternating magnetic flux that generates a measurable ($dc$) potential drop, induced between the inner and the outer edge of the disk \cite{Tomadin}. It also offers new opportunities to experimentally determine the Hall viscosity \cite{Holder-PRL} and the hydrodynamic magnetoresistance that is dominated by the field-induced vorticity of the flow rather than by the field dependence of the kinetic coefficients \cite{Shavit}. Here we elaborate on the latter example focusing on the magnetoresistance in the crossover region of the Gurzhi effect from the Stokes-to-Ohmic flow.   

\begin{figure}
  \centering
   \includegraphics[width=3in]{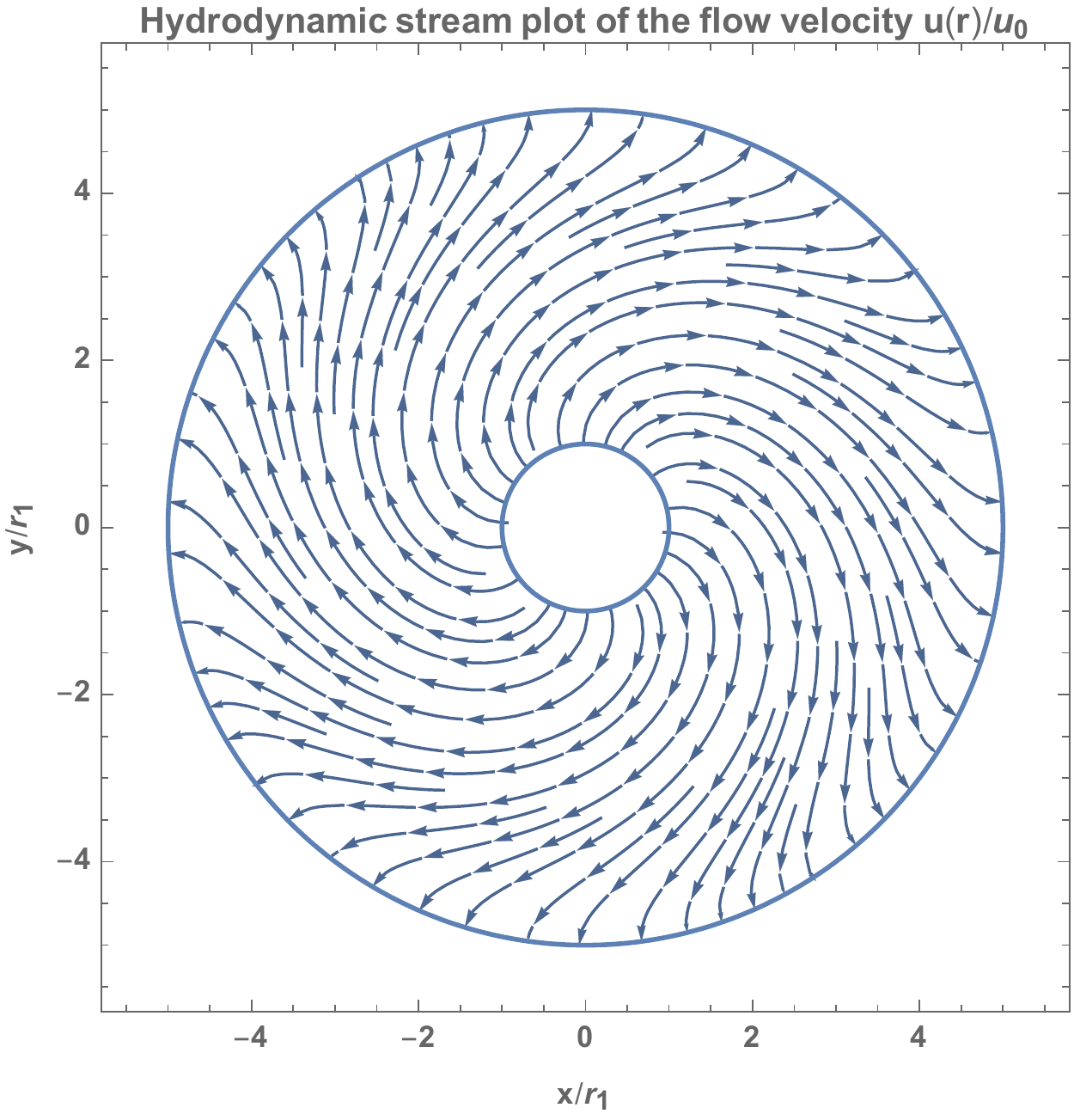}
  \includegraphics[width=3in]{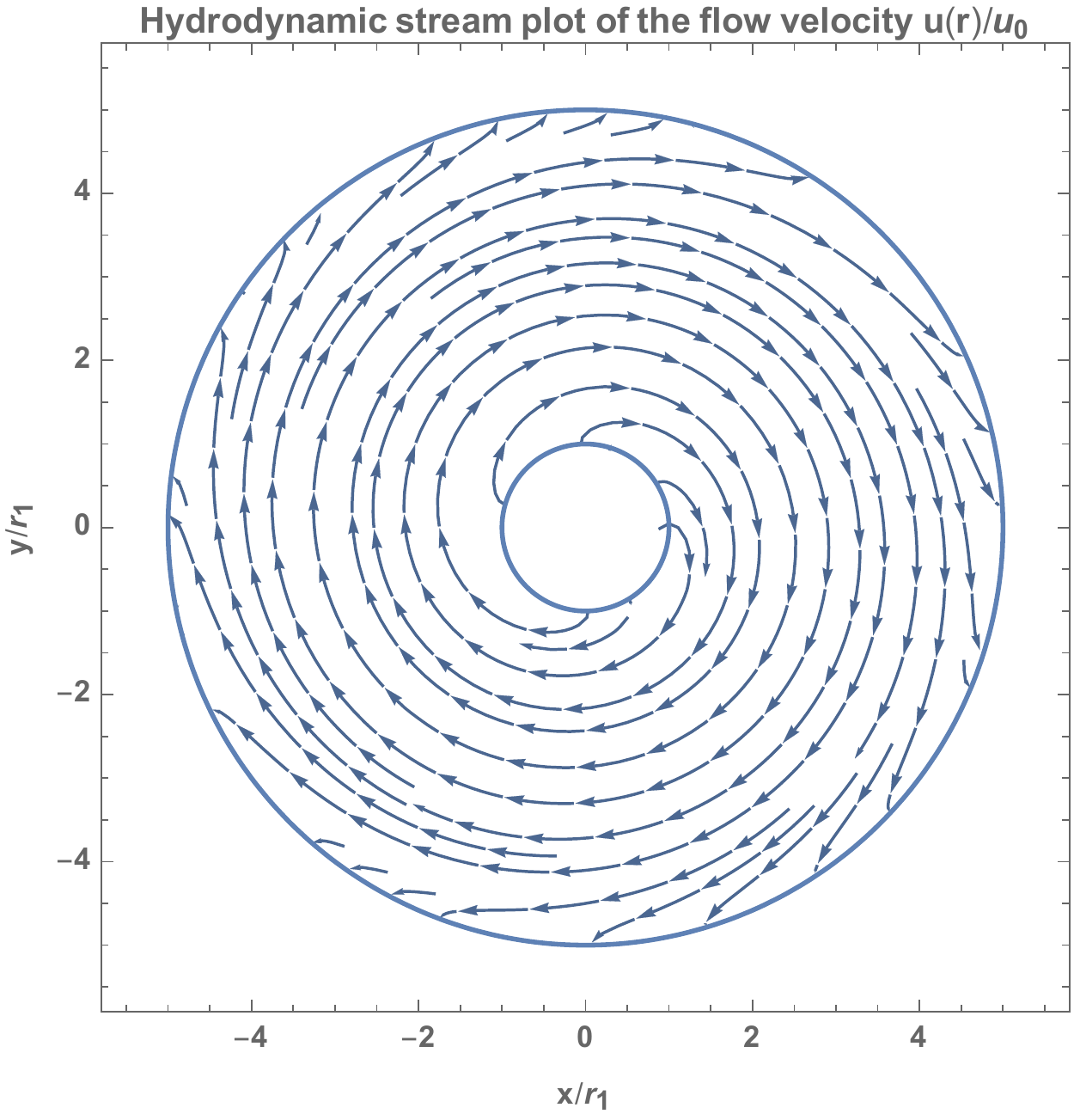}
 \caption{The stream plot of the viscous flow was generated in polar coordinates for $u_x(r)=u_r(r)\cos\phi-u_\phi(r)\sin\phi$ and $u_y(r)=u_r(r)\sin\phi+u_\phi(r)\cos\phi$ based on the solution from Eqs. \eqref{j-r} and \eqref{j-phi}. The velocity field was normalized in units of $u_0=I/(2\pi enr_1)$ for the aspect ratio $a=r_2/r_1=5$. The strength of the external field that controls the distribution of the flow pattern between electrodes in the bulk  is characterized by a dimensionless parameter $q=nr^2_1/(2\eta l^2_B)$, where $l_B=\sqrt{1/eB}$ is the magnetic length. This parameter measures the relative strength of the Lorentz and viscous Stokes forces  and determines the number of turns the flow makes between the electrodes. On the left panel we took $q=0.55$ while on the right panel $q=3.55$ for comparison, the aspect ratio was kept the same in both cases.}  
  \label{Fig-Corbino}
\end{figure}

The centro-symmetry of the Corbino disk suggests the use of polar coordinates. For the purpose of MR calculation we need to project the Navier-Stokes equation \eqref{NS-MHD} into the radial ($u_r$) and azimuthal ($u_\phi$) components of the flow field. The corresponding components of the Laplacian operator are given by \cite{LL-V6}
\begin{equation}
(\nabla^2 \bm{u})_r=\nabla^2 u_r-\frac{u_r}{r^2}-\frac{2}{r^2}\frac{\partial u_\phi}{\partial \phi},\qquad
(\nabla^2 \bm{u})_\phi=\nabla^2 u_\phi-\frac{u_\phi}{r^2}+\frac{2}{r^2}\frac{\partial u_r}{\partial \phi}.
\end{equation}
For an isotropic system with magnetic field perpendicular to the plane of the flow, both components of the flow velocity depend only on the radial coordinate such that terms like $\partial_\phi u_{r,\phi}$ vanish. Consequently, for the corresponding components of the electrical current we find the two equations   
 \begin{equation}
\frac{\eta}{ne}\Delta j_r+enE_r+j_\phi B =\rho_0 nej_r,\quad
\frac{\eta}{ne}\Delta j_\phi+enE_\phi-j_r B =\rho_0 nej_\phi,
 \end{equation} 
where we introduced radial operator $\Delta=\nabla^2-1/r^2$ and expressed the kinematic viscosity $\nu=\eta/(mn)$ in terms of shear viscosity $\eta$. In the current setup there is no azimuthal component of the electric field $E_\phi=0$, but there is a freely circulating current $j_\phi$. The situation here is opposite to that of the Hall bar, with a transversal field but no current. Furthermore, from the continuity equation, current conservation in the radial direction implies 
\begin{equation}\label{j-r}
j_r(r)=I/(2\pi r),
\end{equation} 
which gives an equation for the azimuthal current 
\begin{equation}
(\Delta-\lgu^{-2})j_\phi=\left[\frac{d^2}{dr^2}+\frac{1}{r}\frac{d}{dr}-\left(\frac{1}{r^2}+\frac{1}{\lgu^2}\right)\right] j_\phi=\frac{ne}{\eta}\frac{IB}{2\pi r}.
\end{equation}
This equation coincides with the canonical form of the differential equation for the modified Bessel function 
of the first order, which thus gives us two linearly independent solutions $\mathrm{I}_1(r/\lgu)$ and $\mathrm{K}_1(r/\lgu)$. The special solution due to the right-hand-side can be tried in the form $j_\phi=C(I/r)$ where $C$ is a yet unknown constant. By observing that $\Delta(1/r)=0$ we easily deduce that $C=-B/(2\pi\rho_0 ne)$. As a result, the general solution takes the form    
\begin{equation}\label{j-phi}
j_\phi(r)=\frac{IB}{2\pi\rho_0 ne}\left[C_1\,\mathrm{I}_1(r/\lgu)+C_2\, \mathrm{K}_1(r/\lgu)-\frac{1}{r}\right].
\end{equation}
The integration constants $C_1$ and $C_2$ can be determined from the boundary conditions. For simplicity, we apply no-slip boundary conditions $j_\phi(r_1)=j_\phi(r_2)=0$. To visualize viscous effects we deduced the flow pattern from the obtained solution and plotted $\bm{u}(\bm{r})$ in Fig. \ref{Fig-Corbino}.   

As next step in our analysis we use the components of the stress tensor \cite{LL-V6}
\begin{equation}
\sigma_{rr}=2\eta\frac{\partial u_r}{\partial r}, \quad \sigma_{r\phi}=\eta\left(\frac{1}{r}\frac{\partial u_r}{\partial\phi}+\frac{\partial u_\phi}{\partial r}-\frac{u_\phi}{r}\right), \quad \sigma_{\phi\phi}=2\eta\left(\frac{1}{r}\frac{\partial u_\phi}{\partial\phi}+\frac{u_r}{r}\right), 
\end{equation}
to determine energy dissipation rate due to viscous friction  
 \begin{equation}
W=\frac{1}{2\eta}\sum_{ij}\sigma^2_{ij}dV.
\end{equation}
The latter gives us resistance $R=W/I^2$. As a result we find  
 \begin{equation}
 R=R_0+R_B.
 \end{equation}
The zero field part of the resistance $R_0$ comprises of Ohmic and Stokes contributions. The Ohmic part is determined by the momentum-relaxing scattering time in the bulk of the flow and is given by a standard expression 
\begin{equation}\label{R-Ohm}
R^{\text{Ohm}}_{0}=\frac{\rho_0}{2\pi}\ln(r_2/r_1). 
\end{equation}
This form of the resistance can be readily seen from the Navier-Stokes equation itself by noticing that $\Delta j_r=0$ yields for the radial component of the electric field  $E_r=\rho_0 j_r$, with the corresponding voltage drop $V=\int^{r_2}_{r_1} E_rdr$. This  immediately yields Eq. \eqref{R-Ohm}. The viscous, Stokes contribution to the resistance is given by 
\begin{equation}\label{R-Stokes}
R^{\text{Stokes}}_{0}=\frac{\eta}{\pi (ne)^2} \left(\frac{1}{r^2_1}-\frac{1}{r^2_2}\right),
\end{equation}
but its physical origin is much more subtle and to some extent paradoxical as explained in the recent insightful work \cite{Shavit}. To gauge the relative importance of these two terms one should notice that for the large disk, $r_2\gg r_1$, the viscous term saturates. The Ohmic part, however, grows in this limit very slowly and the ratio between the two is $R^{\text{Ohm}}_0/R^{\text{Stokes}}_{0}\sim (r_1/\lgu)^2\ln(r_2/r_1)$, which means that  the Ohmic part could in principle dominate, even when the Gurzhi length is large. As explained in Ref. \cite{Shavit} the result for $R^{\text{Stokes}}_0$ originates from the voltage drop at the electrodes. In the Ohmic regime, the impact of contact resistance was analyzed in the context of the electronic thermal transport: Lorenz number measurements and Wiedemann-Franz law in particular \cite{KC-Fong}.

\begin{figure}
  \centering
  \includegraphics[width=4.5in]{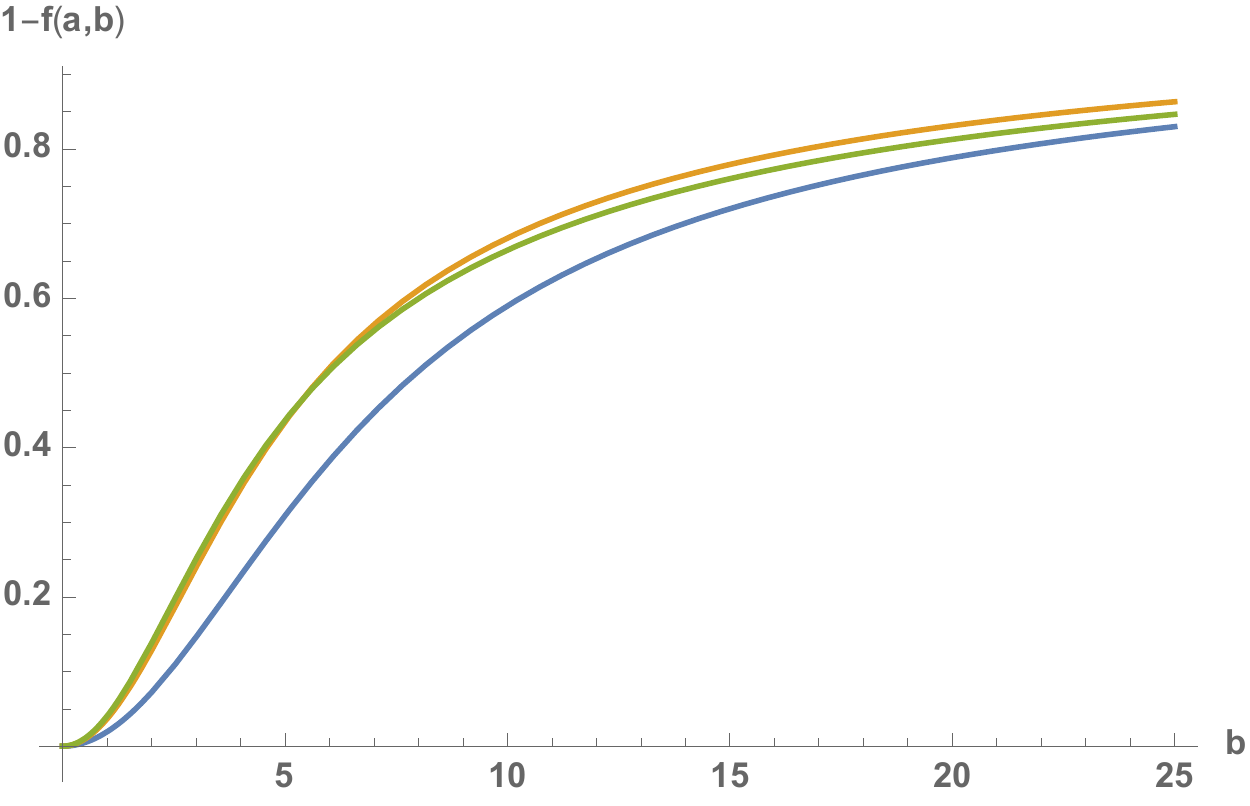}
 \caption{A dimensionless scaling function Eq. \eqref{f-Corbino} that describes Stokes-to-Ohm crossover in the magnetoresistance Eq. \eqref{MR-Corbino} for Corbino device with different choice of aspect ratios $a=2,4,8$.}  
  \label{Fig-crossover}
\end{figure}
  
The field dependent part of the resistance can be presented in the form  
\begin{align}\label{MR-Corbino}
 R_B=\frac{B^2\ln a}{2\pi\rho_0(ne)^2}[1-f(a,b)], \quad a=r_2/r_1, \quad b=r_2/\lgu.
\end{align}
The dimensionless function 
\begin{align}\label{f-Corbino}
f(a,b)=1-\frac{1}{\ln a}\left\{\frac{[\mathrm{I}_0(b)-\mathrm{I}_0(b/a)][(a/b)\mathrm{K}_1(b)-(1/b)\mathrm{K}_1(b/a)]}{\mathrm{I}_1(b/a)\mathrm{K}_1(b)-\mathrm{I}_1(b)\mathrm{K}_1(b/a)}\right. \nonumber\\ 
\left.+\frac{[\mathrm{K}_0(b)-\mathrm{K}_0(b/a)][(a/b)\mathrm{I}_1(b)-(1/b)\mathrm{I}_1(b/a)]}{\mathrm{I}_1(b/a)\mathrm{K}_1(b)-\mathrm{I}_1(b)\mathrm{K}_1(b/a)}\right\}
\end{align}
describes the crossover from the Stokes to the Ohmic regime. This function is plotted in Fig. \ref{Fig-crossover} for several different values of the aspect ratio $a$. Asymptotic limits of this function can be relatively easily extracted. In the Ohmic regime, $b\gg1$, $f$ is a decaying function of $b$ such that to leading order holds: 
\begin{equation}
R^{\text{Ohm}}_B=\frac{B^2\ln a}{2\pi\rho_0 (ne)^2}\propto B^2\tau_{\text{MR}},\quad r_2\gg\lgu.
\end{equation}
In the opposite, viscosity-dominated limit, where $\lgu\gg r_{1,2}$, we can expand the Bessel functions at small argument $ b\ll1$ such that  
\begin{equation}
R^{\text{Stokes}}_B=\frac{B^2r^2_2}{16\pi \eta}\left[1-\frac{1}{a^2}\right]\left[1-\frac{4a^2\ln^2a}{(a^2-1)^2}\right]\propto \frac{B^2}{\tau_{\text{MC}}}\quad r_2\ll\lgu.
\end{equation}   
This result coincides with the earlier conclusion of Refs. \cite{Xie,Shavit} that in the hydrodynamic regime the MR is inversely proportional to the viscosity. This, in principle, enables measurements of the temperature and density dependence of the    viscosity from magneto-transport experiments. On the theoretical side it should be possible to extend these results to cover the ballistic-to-hydrodynamic crossover in the magneto-transport, as was recently done for the geometry of narrow channels \cite{Holder-PRB}. It is also of a special interest to consider magneto-thermo-electric phenomena in Corbino geometry, and Nernst effect in particular \cite{Varlamov-PNAS}. 
  
        
\subsection{Hydrodynamic surface impedance in a viscous skin effect}

In terms of the response to  an electromagnetic field, the hydrodynamic regime of an electron-phonon fluid is not limited to $(dc)$ transport properties but occupies a finite domain of the frequency-momentum $(\omega,q)$ parameter space which is bound by the conditions $\omega\tau_{\mathrm{ep}}\ll1$ and $ql_{\mathrm{ep}}\ll1$. 
Finite frequency properties of viscous electrons have attracted significant theoretical interest in recent years with interesting predictions ranging from nonlinear electrodynamics  \cite{Briskot,Fogler} (e.g. second-harmonic generation), resonant phenomena \cite{Alekseev2} (e.g. viscous cyclotron motion) to nonlocal effects in pulsating flows \cite{Falkovich,Moessner}. The optical conductivity and the transmission of electromagnetic waves through thin ultra-pure metals have been considered in Refs. \cite{Forcella,Toshio} under the condition that hydrodynamic regime is governed by fast electron-electron collisions. Quantum critical hydrodynamics in the $dc$ conductivity of graphene at the neutrality point was predicted in Ref. \cite{Fritz2008} and recently observed experimentally in Ref. \cite{Gallagher2019}. In this section we briefly consider a related problem of the skin-effect (SE) for the strongly coupled electron-phonon liquids.  An observables of interest, discussed in the context of electron hydrodynamics already by Gurzhi in Ref. \cite{Gurzhi-UFN}, is the frequency-dependent surface impedance \cite{Abrikosov}.  

\begin{figure}
  \centering
   \includegraphics[width=3in]{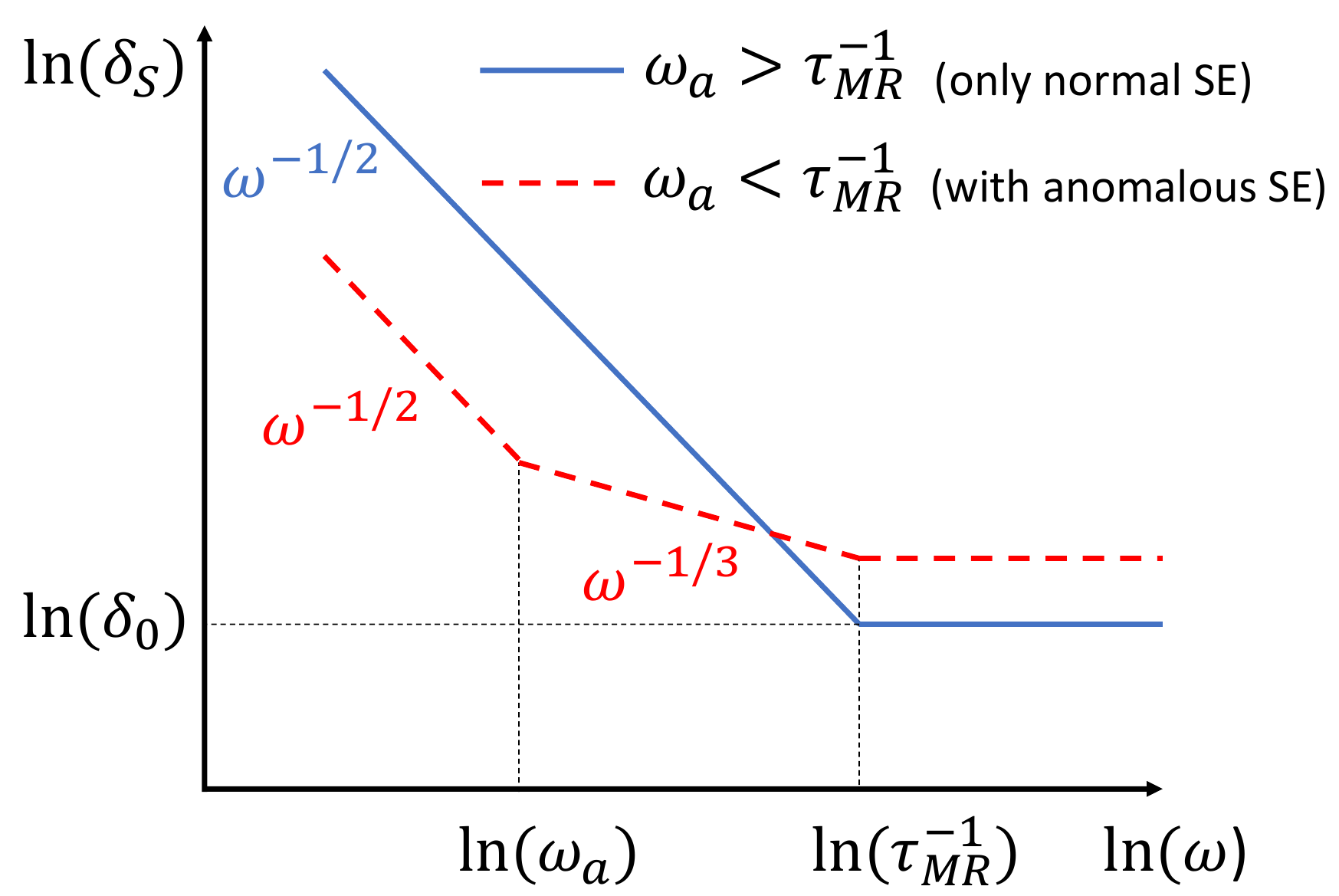}
  \includegraphics[width=3in]{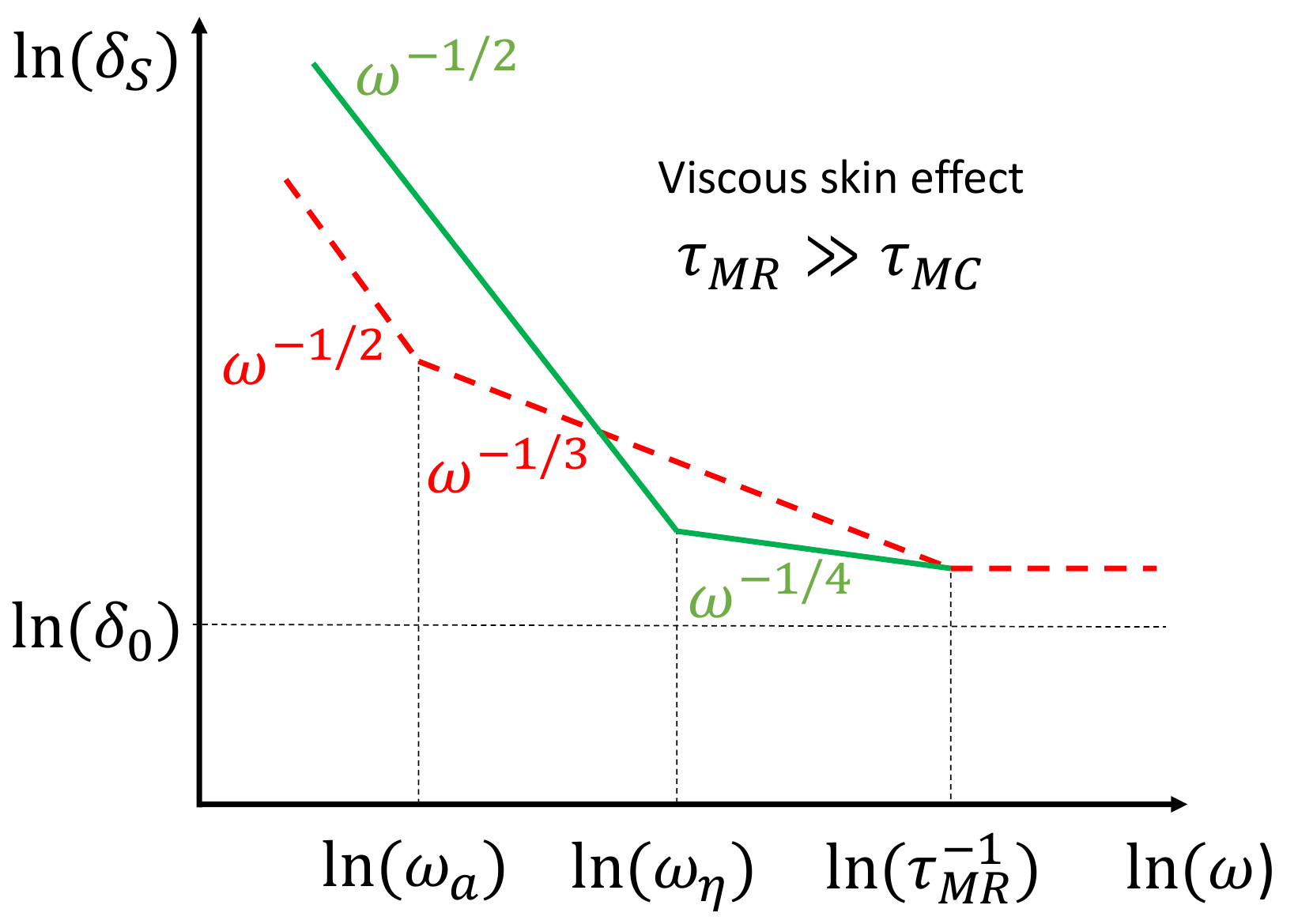}
 \caption{[Left]: Sketch of the frequency dependence of the normal and anomalous skin effect in the log-log scale where $\omega_a$ is the crossover frequency between the normal and anomalous skin effect. One only expects the anomalous skin effect in sufficiently clean samples. On the plot $\delta_0=c/\omega_{\text{pl}}$ and $\omega_a=(c/v_F)^2\tau^{-1}_{\text{MR}}/(\omega_{\text{pl}}\tau_{\text{MR}})^2$ where $\omega_{\text{pl}}$ is the plasma frequency. [Right]: Sketch of the skin effect including the intermediate regime of viscous skin effect behavior with $\omega_\eta\simeq(\tau_{\text{MR}}/\tau_{\text{MC}})\omega_a$}  
  \label{Fig-SE}
\end{figure}

Consider a skin-effect geometry when a monochromatic electromagnetic wave of frequency $\omega$ is incident on a metal surface ($xy$-plane). It is assumed that the metal occupies a semi-infinite volume $z>0$ with the vacuum on the other side $z<0$. From a pair of Maxwell equations 
\begin{equation}
[\nabla\times\bm{E}]=-(1/c)\partial_t\bm{H},\quad [\nabla\times\bm{H}]=(4\pi/c)\bm{j}
\end{equation}
we can establish a self-consistent relation between the electrical field and the induced current in the medium 
\begin{equation}\label{E-j}
\nabla^2\bm{E}=(4\pi/c^2)\partial_t\bm{j}.  
\end{equation}
In the linear regime, the current is proportional to the drift velocity of the liquid 
\begin{equation}
\bm{j}=en\bm{u},
\end{equation}
which obeys our hydrodynamic equation of motion 
\begin{equation}
\partial_t\bm{u}=\nu\nabla^2\bm{u}+e\bm{E}/m-\bm{u}/\tau_{\mathrm{MR}}
\end{equation}
that includes a time-dependent inertia term. By passing to Fourier space in frequency $\bm{E}(\bm{r},t)=\mathrm{Re}\left\{\bm{E}_\omega e^{i\omega t}\right\}$ and eliminating $\bm{u}(\bm{r},t)$, one easily obtains a single linear differential equation for the spatial dependence of the field. For the described geometry one finds  
\begin{equation}\label{E}
\partial^4_zE_\omega-l^{-2}_\mathrm{G}(\omega)\partial^2_zE_\omega+il^{-4}_\omega E_\omega=0.
\end{equation}
Here we introduced the frequency-dependent Gurzhi length
\begin{equation}
l_{\mathrm{G}}(\omega)=l_{\mathrm{G}}/\sqrt{1+i\omega\tau_{\mathrm{MR}}}
\end{equation} 
and also another frequency-dependent length scale 
\begin{equation}
l_\omega=\sqrt[4]{v_F\delta_0^2l_{\mathrm{MC}}/\omega} ,\quad \delta_0^2=mc^2/(4\pi ne^2),
\end{equation}
where $\delta_0$ is the familiar London penetration depth in the clean limit. The surface impedance is defined as the ratio between the electric field on the metal surface and the
current density, integrated over the volume 
\begin{equation}\label{Z}
Z(\omega)=\frac{E_\omega(0)}{(4\pi/c)\int j(z)dz}=-(i\omega/c)[E_\omega(0)/\partial_zE_\omega(0)].
\end{equation}
The impedance is a complex function of frequency and its real part determines the energy dissipated by the field. To find $Z$ we look at the characteristic equation of Eq. \eqref{E}, $E_\omega\propto e^{kz}$, whose roots follow as solutions of a bi-quadratic equation 
\begin{equation}\label{roots}
k^2_\pm(\omega)=\frac{1}{2}\left[l^{-2}_{\mathrm{G}}\pm\sqrt{l^{-4}_{\mathrm{G}}-4il^{-4}_\omega}\right].
\end{equation}
This equation gives four different roots and one needs to select two of them $k_1$ and $k_2$ that have negative real part. These solutions correspond to a decaying field into the bulk of the sample.   
The spatial profile of the field is then given by a linear superposition of two exponentials: $E_\omega(z)=A_0e^{k_1z}+B_0e^{k_2z}$. Two coefficients are determined by the boundary conditions $E_\omega(0)=A_0+B_0$ and $[\partial^3_zE_\omega(z)]_{z=0}=-(1/l_{\mathrm{S}})[\partial^2_zE_\omega(z)]_{z=0}$, where $l_{\mathrm{S}}$ is again the slip length \cite{Kiselev}. The second boundary condition corresponds to the linear relationship between $\bm{E$} and $\bm{u}$ and follows directly from Eqs. \eqref{u-bc} and \eqref{E-j}. Solving the linear algebraic equations we find 
\begin{equation}
A_0=E_\omega(0)\frac{k^2_2}{k^2_2-\beta k^2_1},\quad B_0=-E_\omega(0)\frac{\beta k^2_1}{k^2_2-\beta k^2_1}
\end{equation}
where $\beta=(1+k_1l_{\mathrm{S}})/(1+k_2l_{\mathrm{S}})$. For the case of no-slip boundary condition ($\beta\to1$) one can expresses $Z$ in terms of roots $k_{1,2}$ as follows  
\begin{equation}
Z(\omega)=-\frac{i\omega}{c}\frac{k_1+k_2}{k_1k_2}.
\end{equation}
In the opposite case of no-stress ($\beta\to k_1/k_2$) surface impedance takes the form 
\begin{equation}
Z(\omega)=-\frac{i\omega}{c}\frac{k^2_1+k_1k_2+k^2_2}{k_1k_2(k_1+k_2)}. 
\end{equation}
It turns out that both limits exhibit the same frequency dependence (modulo numerical factors of the order of unity).
 Indeed, there are two special cases of interest that one can analyze. First is the regime when $l_\omega\gg\l_{\mathrm{G}}$, which implies a bound on the range of frequencies $\omega<\omega_\eta$, where $\omega_\eta=\left(\tau_{\rm MR}/\tau_{\rm MC}\right)\omega_a$ is determined by the  frequency $ \omega_a\sim\tau^{-1}_{\mathrm{ep}}(\delta_0/l_{\mathrm{MR}})^2$, where usually the skin effect crosses over to the  anomalous skin effect.  For $\omega<\omega_\eta$ it is easy to see from Eq. \eqref{roots} that one of the roots is parametrically larger than the other: for example $k_1\gg k_2$, with $k_1\sim l^{-1}_{\mathrm{G}}$ and $k_2\sim \delta^{-1}_{\mathrm{S}}$. The length scale $\delta_{\mathrm{S}}=l^2_\omega/l_{\mathrm{G}}=\delta_0/\sqrt{\omega\tau_{\mathrm{MR}}}$ emerges, which is nothing else but the usual skin penetration depth, since $E_\omega(z)\propto e^{-(1+i)z/\sqrt{2}\delta_{\mathrm{S}}}$. The impedance in this frequency range is identical to the one in the normal skin effect  
\begin{equation}
Z(\omega)\approx\frac{\delta_0}{c}\sqrt{\frac{\omega}{\tau_{\mathrm{MR}}}}e^{i\pi/4},\quad \omega<\omega_\eta.
\end{equation}
In the opposite, viscous regime  $\omega> \omega_\eta$ the Gurzhi length is large compared to $l_\omega$. Now there are two parametrically identical roots $k_1=-ik_2=-l^{-1}_\omega e^{-i\pi/8}$ of Eq. \eqref{roots}, and the scale of skin penetration depth is controlled by $l_\omega$ only, such that $\delta_{\mathrm{S}}\propto 1/\sqrt[4]{\omega}$. In this case the impedance is given by  
\begin{equation}
Z(\omega)\approx\frac{\delta_0}{c} \sqrt[4]{\frac{\omega^3v_Fl_{\mathrm{MC}}}{\delta^2_0}}e^{3i\pi/8},\quad \omega_\eta<\omega,
\label{impedance}
\end{equation}
which is solely determined by momentum-conserving electron-phonon collisions. This is the result for no-slip boundary conditions. In the opposite limit, with no stress boundary conditions,   one obtains a  result where $Z(\omega)$ of  Eq. \eqref{impedance} is multiplied by a factor $i/2$. This gives rise to a measurable phase shift in the impedance. Whether no-slip or no-stress boundary conditions are appropriate depends on frequency. The former is correct for $\omega < \omega_\eta \left( l_\mathrm{G}/l_\mathrm{S}\right)^4$ while the latter is appropriate in the opposite limit.  In the regime where the Gurzhi length $l_\mathrm{G}$ is larger than the slip length $l_\mathrm{S}$, which is clearly fulfilled for diffuse scattering at the interface \cite{Kiselev}, this frequency-dependent crossover between distinct  boundary-scattering effects should be observable and may serve as tool to determine the slip length. 

In complete analogy with the Gurzhi effect in the resistance, where the momentum-relaxing scattering rate drops out from the expression for the resistivity, this regime can be  termed as \textit{hydrodynamic skin effect} \cite{Gurzhi-UFN}. The upper bound on frequency that determines the regime of the viscous skin effect is set by the  usual  hydrodynamic condition $l_\omega>l_{\mathrm{MC}}$. It is worth emphasizing that this hydrodynamic limit is conceptually different from the high-frequency anomalous skin effect where $\delta_{\mathrm{S}}\propto 1/\sqrt[3]{\omega}$ and $Z\propto \omega^{2/3}$. Figure \ref{Fig-SE} summarizes the frequency dependence of the surface skin depth in different regimes.  


\subsection{Noise thermometry of electron-phonon scattering}

Johnson noise thermometry provides fruitful experimental tools to study electronic thermoelectric conductivity in solids. Most recently these methods were applied to study electronic conduction of a monolayer graphene over a wide range of temperatures, charge densities, and magnetic fields \cite{Crossno-PhD}. In this section we discuss the role of strong electron-phonon scattering on the noise spectra of current fluctuations in mesoscopic conductors. The question itself is not new and has been discussed by multiple authors employing various approximations and methods of kinetic theory. The comprehensive summary of known results is given in the review article by Blanter and B\"uttiker \cite{BB}, see specifically section 6.3.2 page 122. Perhaps the most concise and elegant summary of work that has been done on this topic is presented in the experimental paper of Steinbach \textit{et al.} \cite{SMD}, see specifically their Fig. 1. To place our approach in the context of existing studies we first briefly summarize key results and acknowledge main contributions. 

The interest in the problem of current noise in mesoscopic conductors was triggered by works of Beenakker and B\"uttiker \cite{CWJBB} based on scattering matrix formalism, and Nagaev \cite{Nagaev-PLA} who employed the stochastic Boltzmann-Langevin kinetic equation (see also book of Kogan \cite{Kogan} on electronic noise and fluctuations in solids for an in-depth overview). These authors showed that the celebrated result of Schottky for a Poisson process of the shot noise, namely the zero-frequency current power spectrum of fluctuations, $S=2eIF$ is suppressed by a Fano factor $F=1/3$. This is a single-particle effect that can be understood from the Dorokhov statistics of transmission eigenvalues in disordered conductors. In the current literature this regime is called  shot noise of cold electrons. The subsequent studies focused on the role of inelastic processes. Frequent electron-electron collisions lead to rapid equilibration. Shot noise survives in this limit and has the same structural form as in the case of noninteracting particles but is described by a different Fano factor $F=\sqrt{3}/4$. This result was demonstrated by Kozub and Rudin \cite{KR}, and de Jong and Beenakker \cite{JCWJB} using a semiclassical approach. These authors assumed that inelastic processes lead to a local equilibrium, described by a Fermi distribution with spatially varying temperature $T(\bm{r})$ and derived an effective diffusion-like equation for the non-equilibrium (voltage-dependent) profile of $T(\bm{r})$. This regime is called shot noise of hot electrons. The crossover between the two and extensions to full-counting statistics was developed by Bagrets \cite{Bagrets} and Gutman \textit{et al}. \cite{Gutman}. The influence of strong electron-phonon collisions was addressed by Gurevich and Rudin \cite{GR}, Nagaev \cite{Nagaev-PRB}, and Naveh \textit{et al.} \cite{NAL}. In the first of these papers the electron-phonon collision integral was treated perturbatively, whereas in the other two papers a two-temperature model of the electron-phonon out-of-equilibrium state was assumed and an equation for  the electronic temperature profile derived. Naveh \cite{Naveh} also performed direct numerical calculations of the integral equation with a phenomenological ansatz for the distribution functions. 

Unlike the calculation of the electron-phonon drag viscosity, where diffusion in momentum space is important, noise is primarily affected by the energy relaxation. For this reason it will be convenient and technically advantageous to average the distribution function over the Fermi surface such that it will depend on the energy and real-space coordinate   
\begin{equation}
n_\varepsilon(\bm{r})=\frac{1}{\nu}\int_{\bm{p}} n_{\bm{p}}(\bm{r})\delta(\varepsilon-\varepsilon_{\bm{p}}),
\end{equation} 
where $\nu$ is the density of states. With this notation, the collision integral due electron-phonon scattering in Eq. \eqref{St-ep} can be rewritten as follows 
\begin{align}
\St_{\text{ep}}\{n,N\}=\int_{\omega} M(\varepsilon,\varepsilon',\omega) [n_{\varepsilon-\omega}(1-n_{\varepsilon})N_{\omega}-n_{\varepsilon}(1-n_{\varepsilon-\omega})(1+N_{\omega})]\nonumber\\
+\int_{\omega} M(\varepsilon,\varepsilon',\omega) [n_{\varepsilon+\omega}(1-n_{\varepsilon})(1+N_{\omega})-n_{\varepsilon}(1-n_{\varepsilon+\omega})N_{\omega}],
\end{align}
where the Eliashberg kernel is of the form 
\begin{equation}
M(\varepsilon,\varepsilon',\omega)=\frac{1}{\nu}\int_{\bm{pq}}W(\bm{p}|\bm{p}'\bm{q})\delta(\varepsilon-\varepsilon_{\bm{p}})\delta(\varepsilon'-\varepsilon_{\bm{p}'})\delta(\omega-\omega_{\bm{q}}).
\end{equation}
Its $\varepsilon,\varepsilon'$ dependence is pinned to energies at the Fermi level, whereas the energy transfer dependence on $\omega$ is strong. Apparently, its functional form in the disordered conductors at frequencies below the scale of Debye energy was subject of certain controversy with multiple conflicting results (this is discussed by Belitz \cite{Belitz}). We will discuss a generic model 
\begin{equation}
M(\omega)=\lambda_{\text{ep}}k(\omega/\omega_D)^k/2,\quad k>1
\end{equation}
and show that main results are only weakly dependent on the exponent $k$. Here we use the same convention for the dimensionless coupling constant of electron-phonon interaction $\lambda_{\text{ep}}$ as introduced below Eq. \eqref{Gammaeppe}. To proceed we regroup terms in the collision integral by separating spontaneous emission contributions, namely pieces independent of the bosonic occupation function, and terms proportional to $N_\omega$. Thus we have 
 \begin{align}
\St_{\text{ep}}\{n,N\}=\int_{\omega} M(\omega) 
\big\{[n_{\varepsilon+\omega}(1-n_{\varepsilon})-n_{\varepsilon}(1-n_{\varepsilon-\omega})]+N_\omega[n_{\varepsilon+\omega}+n_{\varepsilon-\omega}-2n_{\varepsilon}]\big\}.
\end{align}
At this point we apply a Fokker-Planck approximation to this integral operator by expanding fermionic occupation factors over the frequency transfer up to quadratic order   
\begin{equation}
n_{\varepsilon\pm\omega}\approx n_\varepsilon\pm\omega\partial_\varepsilon n_\varepsilon+(\omega^2/2)\partial^2_\varepsilon n_\varepsilon.
\end{equation}
Inserting this back into the collision integral we find 
 \begin{align}\label{St-ep-FP}
\St_{\text{ep}}\{n,N\}\approx \mathcal{A}(1-2n_\varepsilon)\partial_\varepsilon n_\varepsilon+\frac{\mathcal{B}}{2}\partial^2_\varepsilon n_\varepsilon,
\end{align}
where the expansion coefficients are
\begin{equation}
\mathcal{A}=\lambda_{\text{ep}}\int_\omega \omega M(\omega)\simeq a\lambda_{\text{ep}}\omega^2_D,\qquad \mathcal{B}=\lambda_{\text{ep}}\int_\omega \omega^2 M(\omega)(1+N_\omega)\simeq b\lambda_{\text{ep}}\omega^3_D, 
\end{equation}
with $a\sim b$ being model-specific numerical pre-factors of order of unity. In this estimation we assumed $T\ll\omega_D$ so that $N_\omega\ll1$ and cut off the integral at the Debye energy. In general, $\mathcal{B}(\bm{r})$ is weakly coordinate dependent which is ignored in the analysis below. The advantage of the Fokker-Planck approximation is threefold: (i) it is not limited to near-equilibrium problems; (ii) it reduces the collision term to a local differential form; (iii) it preserves the nonlinearity of the collision operator. Curiously, the nonlinearity of Eq. \eqref{St-ep-FP} is of the Burgers type and  known in the theory of nonlinear waves \cite{Burgers,Whitham}.    

Consider a quasi-1D geometry of a wire of length $L$ subject to the voltage bias $V$. In the diffusive approximation, the distribution function obeys the following kinetic equation (see Eq. 221 in Ref. \cite{BB})
\begin{equation}
D\nabla^2 n_{\varepsilon}(x)+\St_{\text{ep}}\{n\}=0
\end{equation}
with the collision term taken from Eq. \eqref{St-ep-FP}. Provided that $n_{\varepsilon}(x)$ is known  the general semiclassical expression for the shot noise power of current fluctuations can be expressed in terms of a non-equilibrium steady-state distribution function as follows:
\begin{equation}\label{noise}
S=\frac{4}{RL}\int^{L/2}_{-L/2}dx\int^{+\infty}_{-\infty} n_\varepsilon(x)[1-n_\varepsilon(x)]d\varepsilon,
\end{equation}
where $R$ is the wire resistance. It will be useful to rescale the coordinate $l=x/L$ and energy $\epsilon=\varepsilon/\omega_D$, and introduce the Thouless energy $E_{\text{Th}}=D/L^2$. In these dimensionless variables it follows 
\begin{equation}
\frac{\partial^2n}{\partial l^2}+\lambda_{\text{ep}}\frac{\omega_D}{E_{\text{Th}}}\left[a(1-2n)\frac{\partial n}{\partial\epsilon}+\frac{b}{2}\frac{\partial^2n}{\partial\epsilon^2}\right]=0.
\end{equation}
This non-linear partial differential equation is of the Burgers type \cite{Burgers,Whitham}, which is a prototypical equation to develop discontinuities such as shock waves. 
Recall that the Fokker-Planck approximation implies strong local equilibration, thus in the current context this means a short relaxation length scale as compared to the wire length $L\gg l_{\text{ep}}$. This practically corresponds to an infinite wire limit. Exploring an analogy to nonlinear waves we can attempt searching for a solution in the form of  a ``propagating soliton" $n_\epsilon(l)\to n(\epsilon-ul)$, where the speed is governed by the voltage, namely $u=eV/\omega_D$. This is also physically justified; we simply assume that  the energy dependence is governed by the local electrochemical potential.
The result reads 
\begin{equation}
n_\varepsilon(x)=\left[\exp\left(\beta(\varepsilon-eVx/L)/\omega_D\right)+1\right]^{-1},\quad \beta^{-1}=\frac{(eV/\omega_D)^2E_{\text{Th}}}{a\lambda_{\text{ep}}\omega_D}+b/2a,  
\end{equation} 
and corresponds to a highly non-thermal state with voltage dependent temperature. This is also the point where, perhaps, the Fokker-Planck approach overlaps with previous approximation, in particular a model with a coordinate and voltage dependent electronic temperature.  From Eq. \eqref{noise} it then follows that the current noise in this regime is described by the voltage-dependent Fano factor 
\begin{equation}
S=2eIF,\quad F\simeq eVE_{\text{Th}}/\lambda_{\text{ep}}\omega^2_D.
\end{equation}
The Fano factor drops as $F\propto 1/L^2$ in this regime that corresponds to a suppression of shot noise by inelastic processes. This is in qualitative agreement with Fig. 1 of Ref. \cite{SMD} in the long $L$ asymptote. It is also in a qualitative agreement with other previous conclusions \cite{NAL,Naveh} albeit obtained under different approximations. 


\section{Summary and outlook}\label{Sec:Summary}

In this work we have considered several examples of hydrodynamic effects that can occur in electron liquids under the condition of strong phonon drag. Electrons and phonons form a combined fluid with an emergent joint drift velocity as hydrodynamic variable. The effect is caused by the fact that the relaxation of the total momentum  $\bm{P}_{\rm el}+\bm{P}_{\rm ph}$ is much slower than the momenta $\bm{P}_{\rm el}$ or $\bm{P}_{\rm ph}$ of electrons or phonons alone. This is guaranteed for clean samples with  weak or kinematically forbidden umklapp scattering processes. We have studied coupled kinetic equations for electrons and phonons, and inferred the effective viscosity of this strongly-coupled transport regime as well as its thermal conductivity. The situation happens to be analogous to the viscous flows in the regime of electron-electron dominated collisions with the only difference that momentum-conserving mean free path  has a different temperature dependence. This difference propagate to numerous observables such as the viscous resistance, the Hall resistance, or the surface impedance. 

While our work was primarily motivated by recent experiments, the delafossite metals PdCoO$_2$ and PtCoO$_2$ studied in Refs. \cite{Daou,Moll,Nandi} and PtSn$_4$ of Refs. \cite{Mun2012,Gooth-PtSn4} in particular, we have not yet tried to tailor this analysis to the case of a multi-band conductors or systems with complex Fermi surfaces. Hydrodynamic transport theory of electron-phonon liquids in 3D Weyl or Dirac semimetals is yet to be fully developed. The first required step towards this direction would be to consider a minimal two-band model of a non-compensated metal. The generalized kinetic scheme has to be developed then for a coupled kinetic equations for electron, holes, and phonons. Another interesting possibility is to consider the possibility of a hydrodynamic regime in Luttinger semimetals \cite{Dumitrescu,Link} with the inclusion of electron-phonon scattering. In addition, in these systems an electron-hole imbalance mode is not restricted so severely like in graphene so that an unusual transport regime is possible. To the best our knowledge, electron-phonon drag of imbalanced liquids has not been addressed in the previous studies.     


\section{Acknowledgments}

We thank Graham Baker, Kamran Behnia, Douglas Bonn, Jennifer Coulter, Cory Dean, Igor Gornyi, Philip Kim, Egor Kiselev, Leonid Levitov, Julia Link, Andrew Lucas, Andrew Mackenzie, Dmitrii Maslov, Alexander Mirlin, Roderich Moessner, Nabhanila Nandi, Boris Narozhny, Thomas Scaffidi, Davide Valentinis, and especially Boris Shklovskii for valuable comments and discussions. This work was supported by the National Science Foundation Grant No. DMR-1653661, the Binational Science Foundation Grant No. 2016317, and the European Commission's Horizon 2020 RISE program Hydrotronics (Grant No. 873028). This work was performed in part at the Max Planck Institute for the Physics of Complex Systems, and the Aspen Center for Physics, which is supported by National Science Foundation Grant No. PHY-1607611. 


\appendix

\section{Appendix}\label{Sec:Appendix}

\subsection{Variational solution of the Boltzmann equation for Bloch-Gr\"uneisen conductivity}\label{App:BG}

In this section we present a  method to solve  the linearized Boltzmann equation, which is based on the variational formulation of the kinetic theory. We begin from Eq. \eqref{BKE-BG} and rewrite it by combining both terms on the right-hand-side together, which gives 
\begin{equation}\label{BKE-BG-sym}
\bm{v_p}\frac{\partial f}{\partial\varepsilon_{\bm{p}}}=\frac{1}{T}\int_{\bm{p}'}D(\bm{p},\bm{p}')[\bm{v}_{\bm{p}'}g_{\bm{p}'}-\bm{v_p}g_{\bm{p}}]\sum_{\sigma=\pm}\delta(\varepsilon_{\bm{p}'}-\varepsilon_{\bm{p}}+\sigma\omega_{\bm{p}-\bm{p}'}),
\end{equation} 
where we took a parametrization of the form  
\begin{equation}
\psi_{\bm{p}}=e\bm{v_p}\bm{E}\frac{g(\varepsilon_{\bm{p}})}{T},
\end{equation}
and, after few algebraic steps, reorganized kernels $K_\pm$ to make the result manifestly symmetric with respect to interchange of momenta. This way we arrived at
\begin{equation}
D(\bm{p},\bm{p}')=D_0|\bm{p}-\bm{p}'|\frac{f(\varepsilon_{\bm{p}})f(\varepsilon_{\bm{p}'})}{|e^{-\varepsilon_{\bm{p}'}/T}-e^{-\varepsilon_{\bm{p}}/T}|}. 
\end{equation} 
It is easy to see that Eq. \eqref{BKE-BG-sym} can be obtained from the variational derivative of the following auxiliary functional 
\begin{equation}\label{BKE-BG-Functional}
Q_P[g]=\frac{1}{4T}\int_{\bm{pp}'}D(\bm{p},\bm{p}')[\bm{v}_{\bm{p}'}g_{\bm{p}'}-\bm{v_p}g_{\bm{p}}]^2\sum_{\sigma=\pm}\delta(\varepsilon_{\bm{p}'}-\varepsilon_{\bm{p}}+\sigma\omega_{\bm{p}-\bm{p}'})-\int_{\bm{p}}\bm{v}^2_{\bm{p}}g_{\bm{p}}\frac{\partial f}{\partial\varepsilon_{\bm{p}}}.
\end{equation}
Thus solving Eq. \eqref{BKE-BG-sym} is fully equivalent to minimizing Eq. \eqref{BKE-BG-Functional}.  Of course,  this is not an easy task either.  However, one can try to postulate a variational ansatz for $g_{\bm{p}}$ and then extremize the functional, which is often a simpler computation. To this end, suppose that $g_{\bm{p}}=g$ is a constant, which is the leading contribution for temperatures small compared to the Fermi energy,  we have
\begin{equation}
Q_P[g]=\frac{1}{2}A_Pg^2-B_Pg,
\end{equation}
where 
\begin{equation}
A_P=\frac{1}{2T}\int_{\bm{pp}'}D(\bm{p},\bm{p}')[\bm{v}_{\bm{p}'}-\bm{v_p}]^2\sum_{\sigma=\pm}\delta(\varepsilon_{\bm{p}'}-\varepsilon_{\bm{p}}+\sigma\omega_{\bm{p}-\bm{p}'}), \quad 
B_P=\int_{\bm{p}}\bm{v}^2_{\bm{p}}\frac{\partial f}{\partial\varepsilon_{\bm{p}}}=v^2_F\nu.
\end{equation}
Here in the integral for $B_P$ we introduced density of states $\nu$ at the Fermi energy. The extremal $Q_P$ is determined by $g=B_P/A_P$. This allows us to determine the conductivity as
\begin{equation}
\sigma_{\alpha\beta}=2e^2\frac{B_P}{A_P}\int_{\bm{p}}\bm{v}_{\bm{p}\alpha}\bm{v}_{\bm{p}\beta}\left(-\frac{\partial f}{\partial\varepsilon_{\bm{p}}}\right)=\frac{2e^2}{3}\frac{(v^2_F\nu)^2}{A_P}\delta_{\alpha\beta}=\sigma_B\delta_{\alpha\beta},
\end{equation}
thus finding temperature dependence of the Bloch-Gr\"uneisen conductivity $\sigma_B(T)$ is reduced to the computation of the $A_P(T)$. For the latter we have 
\begin{align}
&A_P(T)=\frac{v^2_F}{sT}\int d\varepsilon d\varepsilon' d\omega F_P(\varepsilon,\varepsilon',\omega)\frac{\omega f(\varepsilon)f(\varepsilon')}{|e^{-\varepsilon/T}-e^{-\varepsilon'/T}|}\sum_{\sigma=\pm}\delta(\varepsilon'-\varepsilon+\sigma\omega),\\ 
& F_P(\varepsilon,\varepsilon',\omega)=\frac{D_0}{v^2_F}\int_{\bm{pp}'}(\bm{v}_{\bm{p}}-\bm{v}_{\bm{p}'})^2\delta(\omega-\omega_{\bm{p}-\bm{p}'})
\delta(\varepsilon-\varepsilon_{\bm{p}})\delta(\varepsilon'-\varepsilon_{\bm{p}'}).
\end{align}
Since electronic momenta are close to Fermi momentum, and the phonon momentum is small, the following approximations apply: $(\bm{v}_{\bm{p}}-\bm{v}_{\bm{p}'})^2\approx 2v^2_F(1-\cos\theta_{\bm{pp}'})$ and $\omega_{\bm{p}-\bm{p}'}\approx\sqrt{2}sp_F\sqrt{1-\cos\theta_{\bm{pp}'}}$. This implies that to the leading order $F(\varepsilon,\varepsilon',\omega)$ is independent of $\varepsilon,\varepsilon'$ so that 
\begin{equation}
F_P(\varepsilon,\varepsilon',\omega)\approx D_0\nu^2\int^{\pi}_{0}d\theta \sin\theta(1-\cos\theta)\delta(\omega-\sqrt{2}sp_F\sqrt{1-\cos\theta})=
\frac{D_0\nu^2}{2sp_F}\Theta(2sp_F-\omega)\left(\frac{\omega}{sp_F}\right)^3,
\end{equation}
where $\Theta(x)$ is the Heaviside step function. Next, we observe that under the approximation that $F_P$ only depends on $\omega$, the energy integrations in $A_P(T)$ can be performed in the closed form. Indeed, it can be readily verified that 
\begin{equation}\label{integral}
\sum_{\sigma=\pm}
\int d\varepsilon d\varepsilon' \frac{f(\varepsilon)f(\varepsilon')}{|e^{-\varepsilon/T}-e^{-\varepsilon'/T}|}\delta(\varepsilon'-\varepsilon+\sigma\omega)=\frac{\omega}{\cosh(\omega/T)-1}.
\end{equation}
Finally, combining everything together as a result we obtain  with $\omega_D\approx 2sp_F$
\begin{equation}
A_P(T)=16D_0\nu^2v^2_Fp_FG\left(\frac{T}{\omega_D}\right),\quad G(t)=t^5\int^{t^{-1}}_{0}\frac{x^5dx}{\cosh x-1}. 
\end{equation}
As it was done in the main text, we can define the electron-phonon scattering time of momentum relaxation $\tau_1$ via Bloch-Gr\"uneisen conductivity $\sigma_B=e^2n\tau_1/m$ with 
\begin{equation}
\tau^{-1}_{1}=2 \omega_D \lambda_{\text{ep}}G(t)=\left\{\begin{array}{cc}
480\zeta(5)\lambda_{\text{ep}}\frac{T^5}{\omega_{\rm D}^4} & t\ll1 \\ \lambda_{\text{ep}}T & t\gg1
\end{array}\right.
\end{equation}
This is the well-known Bloch-Gr\"uneisen behavior. As we saw, the implicit assumption of the analysis is that
the phonons remain in equilibrium such that the total momentum conservation is violated.


\subsection{Detailed calculation of the electron-phonon drag viscosity}\label{App:Visc}

In the phonon-drag regime, where the total momentum conservation is respected, the conductivity is infinite (without umklapp and impurity scattering), yet the joint electron-phonon fluid has a common flow viscosity. To this end, we analyze the problem
for a finite shear flow with velocity gradient such that
\begin{equation}
T_{xy}=\eta\frac{\partial u_x}{\partial y}. 
\end{equation}
By starting out from the linearized coupled Boltzmann equations 
\begin{equation}
-\frac{\partial f}{\partial\varepsilon_{\bm{p}}}v_{y}p_x\frac{\partial u_x}{\partial y}=\delta\St_{\text{ep}}\{\psi,\phi\},\quad -\frac{\partial b}{\partial\omega_{\bm{q}}}s_yq_x\frac{\partial u_x}{\partial y}=\delta\St_{\text{pe}}\{\psi,\phi\},
\end{equation}
we first solve for the phonon distribution
\begin{equation}
\phi_{\bm{q}}=\frac{1}{\gamma_{\bm{q}}}\frac{\partial b}{\partial\omega_{\bm{q}}}s_yq_x\frac{\partial u_x}{\partial y}-
\frac{1}{2\gamma_{\bm{q}}}\int_{\bm{pp}'}D(\bm{p},\bm{p}')(\psi_{\bm{p}'}-\psi_{\bm{p}})\sum_{\sigma=\pm}\sigma\delta(\varepsilon_{\bm{p}'}-\varepsilon_{\bm{p}}+\sigma\omega_{\bm{q}})\delta_{\bm{p}'-\bm{p}+\sigma\bm{q}}
\end{equation}
where 
\begin{equation}\label{gamma}
\gamma_{\bm{q}}=\frac{1}{2}\sum_{\sigma=\pm}\int_{\bm{pp}'}D(\bm{p},\bm{p}')\delta(\varepsilon_{\bm{p}'}-\varepsilon_{\bm{p}}+\sigma\omega_{\bm{q}})\delta_{\bm{p}'-\bm{q}+\sigma\bm{q}}.
\end{equation}
It holds that $\phi_{\bm{q}}=\phi_{-\bm{q}}$. We can now insert this solution into the expression for the electronic collision operator
and obtain the effective purely electronic Boltzmann equation
\begin{align}\label{BKE-Effective}
&R_{\bm{p}}\frac{\partial u_x}{\partial y}=
\int_{\bm{pp}'}D(\bm{p},\bm{p}')(\psi_{\bm{p}'}-\psi_{\bm{p}})\sum_{\sigma=\pm}\delta(\varepsilon_{\bm{p}'}-\varepsilon_{\bm{p}}+\sigma\omega_{\bm{p}-\bm{p}'})\nonumber\\ 
&-\int_{\bm{kk}'\bm{p}'}\frac{D(\bm{p},\bm{p}')D(\bm{k},\bm{k}')}{2\gamma_{\bm{p}-\bm{p}'}}(\psi_{\bm{p}'}-\psi_{\bm{p}})
\sum_{\sigma\sigma'=\pm}\sigma\sigma'
\delta(\varepsilon_{\bm{p}'}-\varepsilon_{\bm{p}}+\sigma\omega_{\bm{p}-\bm{p}'})
\delta(\varepsilon_{\bm{k}'}-\varepsilon_{\bm{k}}+\sigma\omega_{\bm{p}-\bm{p}'})
\delta_{\bm{p}'-\bm{p}+\bm{k}-\bm{k}'}.
\end{align}
It contains now the renormalized source term
\begin{equation}
R_{\bm{p}}=-\left(\frac{\partial f}{\partial\varepsilon_{\bm{p}}}v_yp_x+\int_{\bm{p}'}\frac{D(\bm{p},\bm{p}')}{\gamma_{\bm{p}-\bm{p}'}}\frac{\partial b(\omega_{\bm{p}-\bm{p}'})}{\partial\omega_{\bm{p}-\bm{p}'}}s_{\bm{p}-\bm{p}',y}(p_x-p'_x)\sum_{\sigma=\pm}
\delta(\varepsilon_{\bm{p}'}-\varepsilon_{\bm{p}}+\sigma\omega_{\bm{p}-\bm{p}'})\right)
\end{equation}
and the collision term captured by the second contribution on the right-hand-side of Eq. \eqref{BKE-Effective}. Let us estimate the second (integral) term of $R_{\bm{p}}$ that we denote in the following as $\delta R_{\bm{p}}$. First we notice that with the help of Eq. \eqref{integral} $\gamma_{\bm{q}}$ defined in Eq. \eqref{gamma} can be reduced to the following form 
\begin{equation}
\gamma_{\bm{q}}=\left(\frac{D_0\nu}{4v_F}\right)\Theta(2k_F-q)\frac{\omega_{\bm{q}}}{\cosh(\omega_{\bm{q}}/T)-1}. 
\end{equation}
Next we notice that due to kinematics $\omega_{\bm{k}-\bm{k}'}=\sqrt{2}sp_F\sqrt{1-\cos\theta_{\bm{kk}'}}\approx sp_F\theta_{\bm{kk}'}$ so that 
\begin{align}
\delta R_{\bm{p}}=-4vp_F\int d\theta \sin\theta [\sin\varphi-\sin(\theta+\varphi)][\cos\varphi-\cos(\theta+\varphi)] \nonumber \\ 
\times[\cosh(\omega_\theta/T)-1]\frac{\partial b(\omega_
\theta)}{\partial\omega_\theta}\sum_{\sigma=\pm}\frac{f(\varepsilon_{\bm{p}})f(\varepsilon_{\bm{p}}-\sigma\omega_\theta)}{|e^{-\varepsilon_{\bm{p}}/T}-e^{-(\varepsilon_{\bm{p}}-\sigma\omega_\theta)/T}|},
\end{align}
where we took $\bm{p}=p_F(\cos\varphi,\sin\varphi)$ and $\bm{p}'=p_F(\cos(\varphi+\theta),\sin(\varphi+\theta))$. The integral is dominated by the small angle of scattering $\theta=\omega/sp_F\ll1$, so that recalling that $e^{\varepsilon/T}f(\varepsilon)=1-f(\varepsilon)$, summing over $\sigma=\pm$, using Eq. \eqref{f-b-relations}, and expanding over $\omega$ to  leading order we get  
\begin{equation}
\delta R_{\bm{p}}\approx4vp_F\sin\varphi\cos\varphi\frac{\partial f}{\partial\varepsilon_{\bm{p}}}\int^{\infty}_{0}\frac{\omega^3d\omega}{(sp_F)^4}[\cosh(\omega/T)-1]\frac{\partial b_\omega}{\partial\omega}\frac{\omega[1+\coth(\omega/2T)]}{2\sinh(\omega/2T)}e^{-\omega/2T}
\end{equation}
which yields 
\begin{equation}
\delta R_{\bm{p}}=-v_xp_y\frac{\partial f}{\partial\varepsilon_{\bm{p}}}\frac{16\pi^4}{15}\left(\frac{T}{sp_F}\right)^4.
\end{equation}
It is clear that at low temperatures we can ignore the second term in $R_{\bm{p}}$ compared to the first one. The primary
mechanism by which the flow gradient couples to the electron-phonon fluid is by directly affecting its electron
component. By the same token one can estimate the renormalization piece of the collision integral, namely the second integral term on the right-hand-side of Eq. \eqref{BKE-Effective}. It happens to be smaller than the first term and can be also dropped. In the end, we arrive at the much simplified Boltzmann equation 
\begin{equation}\label{BKE-Viscosity}
-\frac{\partial f}{\partial\varepsilon_{\bm{p}}}v_yp_x\frac{\partial u_x}{\partial y}=\int_{\bm{p}'}D(\bm{p},\bm{p}')(\psi_{\bm{p}'}-\psi_{\bm{p}})\sum_{\sigma=\pm}\delta(\varepsilon_{\bm{p}'}-\varepsilon_{\bm{p}}+\sigma\omega_{\bm{p}-\bm{p}'}), 
\end{equation} 
which is essentially the Boltzmann equation without taking into account that the phonons are not equilibrated. Hence, momentum conservation, while important for the hydrodynamic interpretation of the viscosity is not important for its actual value. To proceed with the solution of Eq. \eqref{BKE-Viscosity} we can follow an analysis that is essentially the same as the one we used to determine the resistivity within the Bloch-Gr\"uneisen limit. We can in fact perform this analysis for a distribution function $\propto \cos(l\theta)$, where $l$ is the angular momentum. The resistivity corresponds to $l=1$ while the viscosity
to $l=2$. This yields the scattering rate for arbitrary $l$, and for viscosity in particular 
\begin{equation}
\tau^{-1}_{2}=6 \lambda_{{\rm ep}} \frac{T^5}{\omega_D^4}\int_0^{\omega_D/T}\frac{x^2\left(1-\frac{T^2}{2\omega_D^2}x^2\right)}{\cosh(x)-1}dx
\end{equation}
The  asymptotic behavior in the low-temperature regime gives Eq. \eqref{ep-viscosity} in the main text.  


\subsection{Detailed calculation of the electron-phonon drag thermal conductivity}\label{App:ThCond}

For the sake of thermal conductivity calculation we can make the following ansatz for the non-equilibrium distribution function of electrons 
\begin{equation}
\delta n_1=-T\frac{\partial f}{\partial\varepsilon_{\bm{p}}}\psi_{\bm{p}},\quad \psi_{\bm{p}}=(\bm{v}\nabla_{\bm{r}}T)\frac{\varepsilon_{\bm{p}}}{T}\frac{g(\varepsilon_{\bm{p}})}{T}.
\end{equation}
Then, the  Boltzmann equation for $g_{\bm{p}}$   be obtained from the variational analysis of the functional
\begin{equation}
Q_E[g]=\frac{1}{4T}\int_{\bm{pp}'}D(\bm{p},\bm{p}')\left[\bm{v}_{\bm{p}'}g_{\bm{p}'}\frac{\varepsilon_{\bm{p}'}}{T}-\bm{v_p}g_{\bm{p}}\frac{\varepsilon_{\bm{p}}}{T}\right]^2\sum_{\sigma=\pm}\delta(\varepsilon_{\bm{p}'}-\varepsilon_{\bm{p}}+\sigma\omega_{\bm{p}-\bm{p}'})-\int_{\bm{p}}\bm{v}^2_{\bm{p}}g_{\bm{p}}\frac{\varepsilon^2_{\bm{p}}}{T^2}\frac{\partial f}{\partial\varepsilon_{\bm{p}}}.
\end{equation}  
The analysis here parallels with that of Bloch-Gr\"uneisen calculations with the only difference that we have now some extra energy factors as we seek the results for the thermal current in response to applied temperature gradient. At temperatures small compared to the Fermi energy we can assume that $g(\varepsilon_{\bm{p}})=g$ is a constant and we obtain
\begin{equation}
Q_E[g]=\frac{1}{2}A_Eg^2-B_Eg,
\end{equation}
where 
\begin{equation}
A_E=\frac{1}{2T^3}\int_{\bm{pp}'}D(\bm{p},\bm{p}')[\bm{v}_{\bm{p}'}\varepsilon_{\bm{p}'}-\bm{v_p}\varepsilon_{\bm{p}}]^2\sum_{\sigma=\pm}\delta(\varepsilon_{\bm{p}'}-\varepsilon_{\bm{p}}+\sigma\omega_{\bm{p}-\bm{p}'}), \quad 
B_E=\frac{1}{T^2}\int_{\bm{p}}\bm{v}^2_{\bm{p}}\varepsilon^2_{\bm{p}}\frac{\partial f}{\partial\varepsilon_{\bm{p}}}={v^2_F}{T}c_{\text{el}}(T).
\end{equation}
In the analysis of the coefficient $A_E(T)$ we can introduce the corresponding function $F_E(\varepsilon,\varepsilon',\omega)$:
\begin{align}
&A_E(T)=\frac{v^2_F}{sT}\int d\varepsilon d\varepsilon' d\omega F_E(\varepsilon,\varepsilon',\omega)\frac{\omega f(\varepsilon)f(\varepsilon')}{|e^{-\varepsilon/T}-e^{-\varepsilon'/T}|}\sum_{\sigma=\pm}\delta(\varepsilon'-\varepsilon+\sigma\omega),\\ 
&F_E(\varepsilon,\varepsilon',\omega)=\frac{D_0}{T^2v^2_F}\int_{\bm{pp}'}[\bm{v}_{\bm{p}'}\varepsilon_{\bm{p}'}-\bm{v_p}\varepsilon_{\bm{p}}]^2
\delta(\omega-\omega_{\bm{p}-\bm{p}'})
\delta(\varepsilon-\varepsilon_{\bm{p}})\delta(\varepsilon'-\varepsilon_{\bm{p}'}).
\end{align}
Next we notice that 
\begin{equation}
[\bm{v}_{\bm{p}'}\varepsilon_{\bm{p}'}-\bm{v_p}\varepsilon_{\bm{p}}]^2\approx v^2_F(\varepsilon-\varepsilon')^2-2v^2_F\varepsilon\varepsilon'(1-\cos\theta_{\bm{pp}'}), 
\end{equation}
where the second term contains the usual transport scattering cross-section factor $(1-\cos\theta_{\bm{pp}'})$, however unlike in the case of conductivity, here it gives only a subleading correction for the energy relaxation, and can be neglected. As a result one finds 
\begin{equation}
F_E(\varepsilon,\varepsilon',\omega)\approx\frac{4\lambda_{\text{ep}}s}{\omega_D}\left(\frac{\varepsilon-\varepsilon'}{T}\right)^2\left(\frac{\omega}{\omega_D}\right).
\end{equation}
Owing to the energy conserving delta-function in $A_E$ one can replace $(\varepsilon-\varepsilon')^2\to\omega^2$ in the final integrations.
Finally, calculating the energy current from Eq. \eqref{j-E} 
\begin{equation}
\bm{j}_\varepsilon=-\int_{\bm{p}}\bm{v}_{\bm{p}}(\bm{v}_{\bm{p}}\nabla_{\bm{r}}T)\left(\frac{\varepsilon_{\bm{p}}}{T}\right)^2f_{\bm{p}}(1-f_{\bm{p}})g_{\bm{p}}=-\kappa_{\text{ep}}\nabla T,
\end{equation}
with $g_{\bm{p}}=B_E/A_E$, we determine that the time scale $\tau_E$, that defines thermal conductivity $\kappa_{\text{ep}}$ in Eq. \eqref{kappa-ep}, is given by 
\begin{equation}
\tau^{-1}_{E}=\frac{2\lambda_{\text{ep}}}{\omega^2_DT^3}\int^{\omega_D}_{0}\frac{\omega^5d\omega}{\cosh(\omega/T)-1}.
\end{equation}


\subsection{Spectral properties of the collision integral and super-diffusion on a Fermi surface}\label{App:SuperDiff}

In the context of electron liquids when the electron-electron interaction establishes a hydrodynamic regime it is known that there is a fundamental difference between the relaxation of even and odd modes of the distribution function which is specific to the two-dimensional case. As first shown by Gurzhi and coauthors \cite{GKR,GKK} the ratio of corresponding decay rates is $\gamma_{\text{odd}}/\gamma_{\text{even}}\sim (T/E_F)^2\ll1$ and physically comes from the kinematics of head-on collisions. This problem was recently re-analyzed in the work by Ledwith \textit{et al.} \cite{Ledwith} where special attention was paid to the dependence of these rates on the angular momentum. It was found that $(\gamma_l)_{\text{even}}\sim (T^2/E_F)\ln l$ whereas $(\gamma_l)_{\text{odd}}\sim (T^4/E^3_F)l^4\ln l$ for $1<l<l_{\text{max}}\sim \sqrt{E_F/T}$. In the context of graphene with electron-electron Coulomb interaction, it was recently shown that the corresponding rate behaves as
 $\gamma_l \propto (e^2/v_{\rm F})^2 T \left| l \right|$. The non-analytic dependence with respect to the angular mode $l$  gives rise to super-diffusion on the Dirac cone and L\'evy-flight behavior in phase space, described by a Fokker-Planck equation in phase space with a fractional Laplacian \cite{Kiselev2}.
 
It is perhaps surprising, but to the best of our knowledge, a similar analysis has not been carried out for electron-phonon liquids. We are aware of two related studies. In the work by Kabanov and Alexandrov \cite{Kabanov} the lowest eigenmode of the electron-phonon collision operator  corresponding to the energy relaxation was found. This result was obtained by a Fourier transform of the linearized Boltzmann equation that thus can be reduced to an auxiliary problem to an effective Schr\"odinger equation in the P\"oschl-Teller potential. In the work by Gurevich and Laikhtman \cite{Laikhtman} energy and momentum transport in fluids was analyzed in the regime dominated by phonon-phonon collisions. It was shown that at low enough temperatures the relaxation is primarily governed by near-collinear scattering between acoustic phonons. Globally, however, the relaxation is hierarchical. These collisions first thermalize unidirectional modes on fast scale leading to angle-dependent temperature, which is followed by a slower relaxation process of angular diffusion on a 2D sphere in 3D momentum space. Below we present general results for the electron-phonon collisions applicable for any angular harmonic of non-equilibrium distributions and carry out the  analysis for the 3D case where we reveal the super-diffusive character of the relaxation.

We aim to solve the linearized Boltzmann equation 
\begin{equation}
\left(\frac{\partial}{\partial t}+\bm{v}\nabla_{\bm{r}}\right)\left(-T\frac{\partial f}{\partial\varepsilon_{\bm{p}}}\right)\psi_{\bm{p}}(\bm{r},t)=\delta\St_{\text{ep}}\{\psi\}+S_{\bm{p}}
\end{equation}
with the source term $S_{\bm{p}}$, by expanding the non-equilibrium distribution function into angular momentum eigenmodes of the spherical harmonics  
\begin{equation}
\psi_{\bm{p}}=\sum_{lm}Y_{lm}(\theta_{\bm{p}},\varphi_{\bm{p}})\phi_{lm}(\varepsilon_{\bm{p}},\bm{r},t). 
\end{equation}
In the limit of degenerate fermions we can ignore the $\left| \bm{p} \right|$ dependence of $\phi_{lm}$. Then we multiply the Boltzmann equation with the mode expansion by $Y^*_{l'm'}(\theta_{\bm{p}},\varphi_{\bm{p}})$ and integrate over momenta with the usual prescription $\int_{\bm{p}}\to \frac{\nu}{4\pi}\int d\varepsilon_{\bm{p}}\int d\Omega_{\bm{p}}$ where the solid angle measure is $d\Omega_{\bm{p}}=\sin\theta_{\bm{p}}d\theta_{\bm{p}}d\varphi_{\bm{p}}$. Then it follows after the spacetime Fourier transform 
\begin{equation}
(-i\omega+\tau^{-1}_{l})\phi_{lm}\delta_{ll'}\delta_{m,m'}+iv_Fq(a_{lm}\delta_{l',l+1}+b_{lm}\delta_{l',l-1})\phi_{lm}\delta_{m,m'}=S_{lm}
\end{equation}
where 
\begin{equation}
\tau^{-1}_{l}=\int_{\bm{p}}Y_{lm}(\theta_{\bm{p}},\varphi_{\bm{p}})\delta\St\{Y_{lm}\},\quad 
S_{lm}=\int_{\bm{p}}Y_{lm}(\theta_{\bm{p}},\varphi_{\bm{p}})S_{\bm{p}}.
\end{equation}
Here we used that $\tau^{-1}_{l}$ should not depend on $m$ if the system is rotation invariant. The coefficients are $a_{lm}=\sqrt{\frac{(l+1-m)(l+1+m)}{4l(l+2)+3}}$ and $b_{lm}=\sqrt{\frac{(l-m)(l+m)}{4l^2-1}}$. By using the explicit form of the collision integral, the decay rates (inverse relaxation times) for the given angular harmonic can be presented as follows 
\begin{equation}
\gamma_l=\tau^{-1}_{l}=\frac{1}{2(2l+1)}\sum_m\int_{\bm{pp}'}D(\bm{p},\bm{p}')[Y_{lm}(\theta_{\bm{p}},\varphi_{\bm{p}})-Y_{lm}(\theta_{\bm{p}'},\varphi_{\bm{p}'})]^2\sum_{\sigma=\pm}\delta(\varepsilon_{\bm{p}}-\varepsilon_{\bm{p}'}+\sigma\omega_{\bm{p}-\bm{p}'}).
\end{equation} 
The summation over the azimuthal components of the angular momentum can be completed explicitly with the help of the well known formula from the theory of spherical functions 
\begin{equation}
\frac{1}{2l+1}\sum_mY_{lm}(\theta_{\bm{p}},\varphi_{\bm{p}})Y_{lm}(\theta_{\bm{p}'},\varphi_{\bm{p}'})=\frac{1}{4\pi} P_l(\cos\theta_{\bm{p}-\bm{p}'}),
\end{equation}
where $P_l(x)$ are the Legendre polynomials. This leads to the following result for the matrix elements of the collision operator as function of angular momentum:
\begin{align}
&\gamma_l=\frac{1}{2s}\int d\varepsilon d\varepsilon' d\omega F_l(\varepsilon,\varepsilon',\omega)\frac{\omega f(\varepsilon)f(\varepsilon')}{|e^{-\varepsilon/T}-e^{-\varepsilon'/T}|}\sum_{\sigma=\pm}\delta(\varepsilon'-\varepsilon+\sigma\omega),\\ 
 &F_l(\varepsilon,\varepsilon',\omega)=\frac{D_0}{2\pi}\int_{\bm{pp}'}[1-P_l(\cos\theta_{\bm{p}-\bm{p}'})]\delta(\varepsilon-\varepsilon_{\bm{p}})
 \delta(\varepsilon'-\varepsilon_{\bm{p}'})\delta(\omega-\omega_{\bm{p}-\bm{p}'}).
\end{align}
Adopting the same reasoning as explained in Sec. \ref{App:BG}, we can ignore the $\varepsilon,\varepsilon'$ dependency of $F_l$ for small fermionic energies. The result then simplifies considerably and gives for $\omega<\omega_{D}$
\begin{equation}
F_l\left(\omega\right)=\frac{4\lambda_{\text{ep}}s}{\omega_{D}}\left(\frac{\omega}{\omega_{D}}\right)\left(1-P_{l}\left(1-\left(\frac{\omega}{\omega_{D}}\right)^{2}\right)\right),
\end{equation}
with the same convention for the electron-phonon coupling constant $\lambda_{\text{ep}}$ as used earlier. This
yields
\begin{equation}
\gamma_{l}=\frac{2\lambda_{\text{ep}} T^{3}}{\omega_{D}^{2}}\int_{0}^{\frac{\omega_{D}}{T}}dx\frac{x^{3}\left(1-P_{l}\left(1-\left(\frac{T}{\omega_{D}}\right)^{2}x^{2}\right)\right)}{\cosh\left(x\right)-1}.
\end{equation}
For $l=1$ and $l=$2 we recover, of course, the known results for
the scattering rates relevant in the resistivity without drag
\begin{equation}
\gamma_{1}=\left\{ \begin{array}{cc}
480\zeta\left(5\right)\lambda_{\text{ep}}T^{5}/\omega_{D}^{4} & \:\,{\rm if}\,\,T\ll\omega_{D}\\
\lambda_{\text{ep}} T & \:\,{\rm if}\,\,T\gg\omega_{D}
\end{array}\right. ,
\end{equation}
 and for the viscosity 
\begin{equation}
\gamma_{2}=\left\{ \begin{array}{cc}
3\gamma_{1} & \:\,{\rm if}\,\,T\ll\omega_{D}\\
2\gamma_{1} & \:\,{\rm if}\,\,T\gg\omega_{D}
\end{array}\right. .
\end{equation}
To analyze the rate for arbitrary $l$ we first use $1-P_{l}\left(1-y^{2}\right)\approx\frac{1}{2}l\left(l+1\right)y^{2}\left(1+{\cal O}\left(l^{2}y^{2}\right)\right)$.
This expansion is sufficient for temperatures $T\ll\omega_{D}/l$
and yields after a few steps $\gamma_{l}=\frac{1}{2}l\left(l+1\right)\gamma_{1}$.
The situation is more subtle in the regime $\omega_{D}/l\ll T\ll\omega_{D}$.
To analyze the large-$l$ behavior we split $\gamma_{l}=\delta\gamma_{0}-\delta\gamma_{l}$
where 
\begin{equation}
\delta\gamma_{l}=\frac{2\lambda_{\text{ep}} T^{3}}{\omega_{D}^{2}}\int_{0}^{\frac{\omega_{D}}{T}}dx\frac{x^{3}P_{l}\left(1-\left(\frac{T}{\omega_{D}}\right)^{2}x^{2}\right)}{\cosh\left(x\right)-1}.
\end{equation}
Notice that $\delta\gamma_{l}<\delta\gamma_{0} $ for $l\geq 1$.
Next, we employ the identity
\begin{equation}
\sum_{l=0}^{\infty}P_{l}\left(1-x^{2}\right)t^{l}=\frac{1}{\sqrt{\left(t-1\right)^{2}+2tx^{2}}}
\end{equation}
and obtain for the generating function 
\begin{eqnarray}
\delta\gamma\left(t\right) =  \sum_{l=0}^{\infty}\delta\gamma_{l}t^{l} 
  =  \frac{2\lambda_{\text{ep}} T^{3}}{\omega_{D}^{2}}\int_{0}^{\frac{\omega_{D}}{T}}dx\frac{x^{3}}{\sqrt{\left(t-1\right)^{2}+2t\left(\frac{T}{\omega_{D}}\right)^{2}x^{2}}\left(\cosh\left(x\right)-1\right)}.
\end{eqnarray}
The behavior of $\delta\gamma\left(t\rightarrow1-0^{+}\right)$ determines
the large-$l$ asymptotics of $\delta\gamma_{l}$. The generating
function has a well defined limit as $t\rightarrow1$ with leading
corrections that are linear in $1-t$. This implies that $\delta\gamma_{l}$
cannot decay slower than $l^{-2}$. Hence in the regime $T\ll\omega_{D}/l$
follows that 
\begin{equation}
\gamma_{l\rightarrow\infty}=\delta\gamma_{0}=\left\{ \begin{array}{cc}
24\zeta\left(3\right)\lambda_{\text{ep}}T^{3}/\omega_{D}^{2} & \:\,{\rm if}\,\,T\ll\omega_{D}\\
2\lambda_{\text{ep}} T & \:\,{\rm if}\,\,T\gg\omega_{D}
\end{array}\right.
\end{equation}
This analysis reveals that the behavior at low temperatures and for a sufficiently small angular momentum modes $l$ can be captured via diffusion processes on the Fermi surface. However, at any finite $T$,  high angular modes with angular momentum $l\gg \omega_D/T$ undergo super-diffusion.  These results are further discussed in section \ref{subsec_superdiffusion} in the main text.
  


\end{document}